\documentclass{elsartL}
\usepackage{graphicx}
\usepackage{amssymb}
\usepackage{amsmath}
\usepackage{physics}
\usepackage{mathrsfs}
\usepackage{mathtools}
\usepackage{appendix}
\newcommand{\avg}[1]{\left< #1 \right>} 
\begin{document}

\begin{frontmatter}
\title{A brief history of the pion-nucleon coupling constant}
\author{Evangelos Matsinos}

\begin{abstract}
This work provides a brief history of determinations of the pion-nucleon ($\pi N$) coupling constant from $\pi N$ and $N N$ data. From robust analyses of twenty reported values of the charged-pion coupling constant, 
exhibiting sizeable fluctuation, the result $f_c^2 = 762.9^{+6.5}_{-6.2} \cdot 10^{-4}$ is obtained. Similar values are extracted for the other two $\pi N$ coupling constants, $f_0^2$ and $f_p^2$, from fewer data. The 
average values of the various $\pi N$ coupling constants, extracted in this work, suggest no splitting, in agreement with the thesis of the Nijmegen group. Additional analysis of the $f_c^2$ and $f_0^2$ values, both 
reported in four studies, turned to be inconclusive: one of these studies suggests that $f_0<f_c$, whereas another slightly favours $f_0>f_c$; no significant splitting effects are observed in the other two studies. The 
analysis of the low-energy $\pi N$ data with the ETH model indicates significant splitting and, under certain conditions, it implies that $f_0>f_c$. Also discussed in the paper are the electromagnetic corrections, which 
need to be applied to the strong shift and to the total decay width of the ground state of pionic hydrogen in order that estimates for the hadronic $s$-wave $\pi N$ scattering lengths be obtained; this is a relevant 
subject as $f_c^2$ may be extracted from the isovector scattering length by use of the Goldberger-Miyazawa-Oehme sum rule. Regarding the removal of the electromagnetic effects in the $\pi N$ system at threshold, my opinion 
is that Theory must find a way to provide reliable and accurate corrections, matching the level of accuracy of the experimental results.\\
\noindent {\it PACS 2010:} 13.75.Cs; 13.75.Gx; 25.40.Cm; 25.40.Dn; 25.40.Kv; 25.80.Dj; 25.80.Gn; 11.30.-j
%
%
\end{abstract}
\end{frontmatter}

\section{\label{sec:Introduction}Introduction}

That the pion-nucleon ($\pi N$) coupling constant is fundamental in our understanding of the Cosmos has been adequately emphasised in numerous works. In meson-exchange models of the strong interaction, a significantly weaker 
coupling between the pions and the nucleons would have prevented the neutrons from combining fast with protons in the early Universe; they would have decayed before they had any chance to be enmeshed first in deuterons, then 
in other light nuclei. According to the Big-Bang Nucleosynthesis, within half an hour of the Big Bang, all existing matter had assumed the form of free electrons, protons, and helium nuclei (as well as traces of other nuclei 
up to $^7 {\rm Be}$). On the contrary, a significantly stronger coupling would have resulted in the rapid creation of bound diprotons and would have led to a helium-dominated Universe. It is hard to imagine how life could 
emerge in such a Universe: typical stars burn hydrogen to helium for about $90$ \% of their lives. Evidently, the stellar evolution in a helium-dominated Universe would have been greatly contracted. Apart from the obvious 
cosmological implications, the $\pi N$ coupling constant enters a variety of hadronic phenomena, low-energy theorems, useful relations (e.g., the Goldberger-Treiman relation), etc.

Hypothesised as the carrier of the nuclear force by Yukawa in 1935, the pion was discovered in 1947 by means of the - revolutionary at that time - photographic emulsion technique \cite{lattes1947}. Two pion-related Nobel 
Prizes were awarded in successive years: to Yukawa in 1949 ``for his prediction of the existence of mesons on the basis of theoretical work on nuclear forces'' and to Powell in 1950 ``for his development of the photographic 
method of studying nuclear processes and his discoveries regarding mesons made with this method.''

The efforts to determine the value of the coupling constant between pions and nucleons date back to almost the time of the discovery of the pion. My aim in this paper is to provide a brief history of determinations of this 
coupling constant from $\pi N$ and $N N$ data. Each of the values, accepted for analysis in this work, fulfils the following selection criteria.
\begin{itemize}
\item The value had been accompanied by a meaningful uncertainty.
\item The value had been a new result; excluded from this paper are averages appearing in compilations of physical constants or in review works dedicated to the $\pi N$ coupling constant.
\item The value had been `final' within a given methodology, employed in one research programme. I believe that it makes no sense to list and/or analyse `progress' values, i.e., those which are routinely obtained during the 
development phase of each programme.
\item The value had appeared in a peer-reviewed journal. Values, reported in unpublished works, may occasionally be quoted, but they will not be included in the statistical analyses pursued in Sections 
\ref{sec:DeterminationsOfChargedCoupling}, \ref{sec:DeterminationsOfNeutralCoupling}, and \ref{sec:DeterminationsOfOtherCouplings}.
\item No definite proof exists that the value is not correct.
\end{itemize}

Since 1990, when I became acquainted with Pion Physics, I have come across papers providing lists of values of the $\pi N$ coupling constant(s) and obtaining recommended averages from these values. I will make no effort to 
include in this work any of these papers. It makes no sense to add (at least) ten papers to an already long reference list, in particular as I have no intention to include/use here any results from those works.

I start with some useful definitions in Section \ref{sec:Definitions}. The determinations of the values of the various $\pi N$ coupling constants are discussed in Section \ref{sec:Determinations}. That section is split into 
three parts: the first part discusses the early determinations, up and including 1980 (the original title of that section was `Pre-meson-factory determinations'); the second part deals with the determinations between 1981 
and 1997, i.e., the year in which the $7^{\rm th}$ MENU Conference took place - a decisive moment in Hadronic Physics as the validity of the `canonical value' (details will be given later on), which had routinely been 
imported into many studies for nearly two decades, was openly challenged; the last part of Section \ref{sec:Determinations} discusses the determinations after MENU'97, many of which were based on the measurements obtained 
from pionic hydrogen (and deuterium) at the $\pi N$ threshold (vanishing kinetic energy of the incident pion). On the basis of the reported values of Section \ref{sec:Determinations}, I obtain (what I believe to be) meaningful 
averages in Section \ref{sec:DeterminationsOfAverages}, first for the charged-pion coupling constant (most of the reported values relate to this quantity), then for the neutral-pion coupling constant. The values, extracted 
from low-energy $\pi N$ data with the ETH model, are discussed in Section \ref{sec:ETH}; only one value from that research programme is included in Section \ref{sec:DeterminationsOfChargedCoupling}. The main findings of the 
work are summarised in the last section of the paper. Appendix \ref{App:AppA} concerns the corrections, which need to be applied to the two $s$-wave $\pi N$ scattering lengths, in order that the effects of electromagnetic 
(EM) origin be removed; this is a relevant subject because the $\pi N$ coupling constant may be obtained from the total decay width of the ground state of pionic hydrogen~\footnote{A two-parameter fit to the measurements 
of the strong shifts of the $1 s$ states in pionic hydrogen and deuterium, as well as of the total decay width of the ground state of pionic hydrogen, yields a more accurate estimate for the isovector $s$-wave $\pi N$ 
scattering length.} by use of the Goldberger-Miyazawa-Oehme (GMO) sum rule, which is discussed in Appendix \ref{App:AppB}.

\section{\label{sec:Definitions}Definitions}

\subsection{\label{sec:CouplingConstants}The various $\pi N$ coupling constants}

The general form of the $\pi N N$ interaction Lagrangian density involves both pseudoscalar and pseudovector vertices:
\begin{equation} \label{eq:EQ01}
\Delta \mathscr{L}_{\pi N} = - \frac{1}{1+x} \bar{\psi} \gamma^5 \vec{\tau} \cdot \left( g_{\pi N N} i x \vec{\pi} + \frac{f_{\pi N N} \sqrt{4 \pi}}{m_c} \gamma^\mu \partial_\mu \vec{\pi} \right) \psi \, \, \, ,
\end{equation}
where $\vec{\pi}$ and $\psi$ respectively stand for the quantum fields of the pion and of the nucleon, $m_c$ for the mass of the charged pion, and $\vec{\tau}/2$ for the isospin operator of the nucleon. The quantities 
$\gamma^{\mu}$ ($\mu=0, 1, 2, 3$) are the Dirac $4 \times 4$ matrices, satisfying the relation $\{ \gamma^{\mu}, \gamma^{\nu} \}=2 g^{\mu \nu} I_4$, and $\gamma^5=i \gamma^0 \gamma^1 \gamma^2 \gamma^3$.

The quantity $g_{\pi N N}$ in Eq.~(\ref{eq:EQ01}) is known as `pseudoscalar coupling', whereas $f_{\pi N N}$ is the `pseudovector coupling'. The parameter $x$ determines the strength of the pseudoscalar admixture in the 
$\pi N N$ vertex. Both pure pseudovector ($x=0$) and pure pseudoscalar ($x \to \infty$) couplings have been used in the past.

The two couplings of Eq.~(\ref{eq:EQ01}) are linked via the equivalence relation:
\begin{equation} \label{eq:EQ02}
f_{\pi N N}^2 = \left( \frac{m_c}{m_1 + m_2} \right)^2 \frac{g_{\pi N N}^2}{4 \pi} \, \, \, ,
\end{equation}
where $m_1$ and $m_2$ stand for the masses of the two nucleons involved in the $\pi N N$ vertex: $m_1$ of the incoming (incident, initial-state) nucleon, $m_2$ of the outgoing (emitted, final-state) nucleon. (Some authors 
define the two coupling constants differently, e.g., using the transformation $g_{\pi N N} \to g_{\pi N N} \sqrt{4 \pi}$ or $f_{\pi N N} \sqrt{4 \pi} \to f_{\pi N N}$ in Eqs.~(\ref{eq:EQ01},\ref{eq:EQ02}).) The squares of 
these two coupling constants are usually reported. The pseudoscalar coupling $g_{\pi N N}^2$ has also appeared as $g_{\pi N}^2$ or simply $g^2$. Similarly, the pseudovector coupling $f_{\pi N N}^2$ has also appeared as 
$f_{\pi N}^2$ or simply $f^2$. I will use $g^2$ and $f^2$ in this work~\footnote{When addressing issues of the ETH model, I will use $g_{\pi N N}$. This hadronic model contains another coupling constant, $g_{\pi N \Delta}$; 
to discriminate between the two couplings, it is customary to use the full vertex as subscript.}, identifying them with $g_{\pi N N}^2$ and $f_{\pi N N}^2$ of Eq.~(\ref{eq:EQ02}), respectively. As citing both $g_{\pi N N}^2$ 
and $f_{\pi N N}^2$ values would be impractical, Eq.~(\ref{eq:EQ02}) will be used, to transform $g^2$ values into $f^2$ results. The first question one may pose is: Is only one $f^2$ value to be used in all vertices 
involving one pion and two nucleons? If the isospin invariance is broken in the $\pi N$ interaction, the answer to this question is negative.

\begin{figure}
\begin{center}
\includegraphics [width=15.5cm] {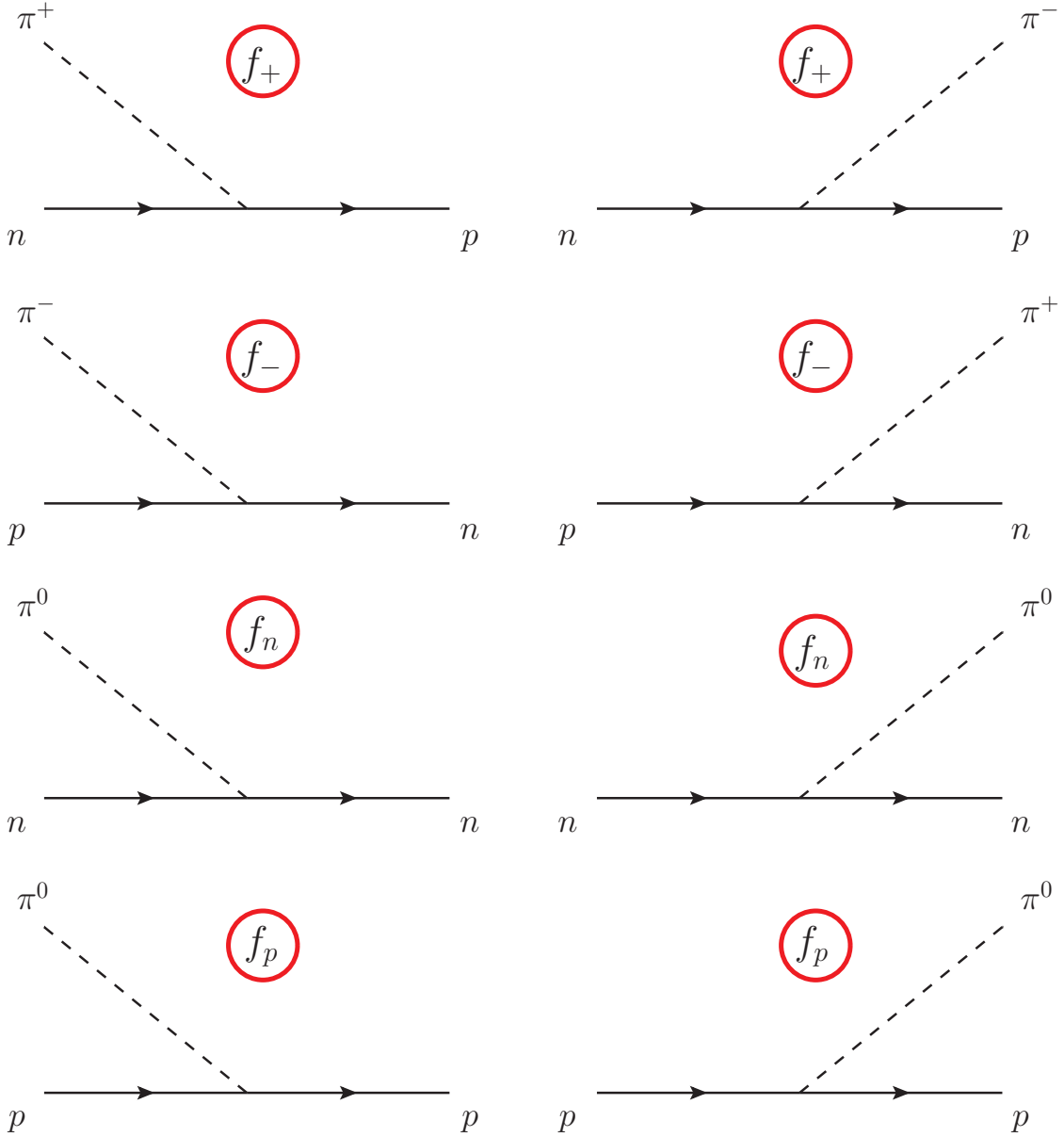}
\caption{\label{fig:CouplingConstants}The general coupling constants between pions and nucleons. The vertices $(\pi^+ n, p)$ and $(n, \pi^- p)$ - where the first elements indicate incoming and the second outgoing particles - 
involve the coupling constant $\sqrt{2} f_+$. The vertices $(\pi^- p, n)$ and $(p, \pi^+ n)$ involve the coupling constant $\sqrt{2} f_-$. The vertices $(\pi^0 n, n)$ and $(n, \pi^0 n)$ involve the coupling constant $-f_n$. 
The vertices $(\pi^0 p, p)$ and $(p, \pi^0 p)$ involve the coupling constant $f_p$. The coefficients and the signs have been chosen in such a way as to result in the equality of all couplings, i.e., $f_+=f_-=f_n=f_p=f$, if 
the isospin invariance is fulfilled \cite{swart1997}.}
\vspace{0.35cm}
\end{center}
\end{figure}

Four coupling constants have been introduced to regulate the strength of the coupling in the various $\pi N$ vertices, see Fig.~\ref{fig:CouplingConstants}: the first one ($f_+$) is associated with the transitions $\pi^+ n \to p$ 
and $n \to \pi^- p$, the second ($f_-$) with $\pi^- p \to n$ and $p \to \pi^+ n$, whereas the remaining two enter the interactions of the neutral pion with the proton ($f_p$) and with the neutron ($f_n$), regardless of 
whether the $\pi^0$ is incoming or outgoing. As a result, the analyses of $\pi^\pm p$ elastic-scattering (ES) measurements determine the product $f_+ f_-$, which is usually denoted as $f_c^2$ or $f_\pm^2$, and is known as 
charged-pion coupling constant. The product $f_p f_n$ enters the description of the $n p$ scattering data ($f_c^2$ is also involved here): $f_0^2 = f_p f_n$, which is known as neutral-pion coupling constant. The analysis 
of the $p p$ data determines $f_p^2$. The early studies had been carried out with only one constant (mostly $f_c$, denoted in those early works simply as $f$). Modern analyses distinguish between $f_c$ and $f_0$, and some 
even determine all three coupling constants: $f_c$, $f_p$, and $f_0$, depending on which databases (henceforth, DBs) are used as input.

\subsection{\label{sec:ScatteringLengths}The $s$-wave $\pi N$ scattering lengths}

Three $s$-wave $\pi N$ scattering lengths (simply `scattering lengths' from now on) are defined as appropriate limits of the scattering amplitudes $\mathcal{F}$ associated with the three experimentally-accessible low-energy 
$\pi N$ reactions: the two ES reactions $\pi^\pm p \to \pi^\pm p$ and the charge-exchange (CX) reaction $\pi^- p \to \pi^0 n$. Assuming no inelasticities (which is a good approximation at low energy), each of these 
scattering amplitudes may be put in the form
\begin{equation*}
\mathcal{F} = \frac{e^{i \, \delta (q)} \sin (\delta (q)) }{q} \, \, \, ,
\end{equation*}
where the quantity $\delta$ is known as the (energy-dependent) phase shift and $q$ denotes the magnitude of the $3$-momentum of the incident pion in the centre-of-momentum (CM) coordinate system. The scattering length $a$ 
is defined as follows.
\begin{equation*}
a = \lim_{q \to 0} \frac{\delta (q)}{q}
\end{equation*}
The three scattering lengths, corresponding to the low-energy $\pi N$ reactions, will be denoted as: $a_{\pi^+ p}$ (for the $\pi^+ p$ reaction), $a_{cc}$ (for the $\pi^- p$ ES reaction), and $a_{c0}$ (for the $\pi^- p$ CX 
reaction); in the context of this work, only $a_{cc}$ and $a_{c0}$ are relevant. Effects of EM origin are present in $a_{\pi^+ p}$, $a_{cc}$, and $a_{c0}$. The removal of these contributions leads to the \emph{hadronic} 
scattering lengths, which will be denoted here as $\tilde{a}_{\pi^+ p}$, $\tilde{a}_{cc}$, and $\tilde{a}_{c0}$.

The fulfilment of the isospin invariance in the $\pi N$ system implies that the three scattering lengths $\tilde{a}_{\pi^+ p}$, $\tilde{a}_{cc}$, and $\tilde{a}_{c0}$ may be expressed as suitable combinations of two quantities, 
i.e., of the scattering lengths in isospin (I) basis: $a_3$ for $I=3/2$ and $a_1$ for $I=1/2$. (No tilde is placed over $a_3$ and $a_1$, as these quantities are defined within the context of the isospin invariance in the 
$\pi N$ interaction.) The relations are: $\tilde{a}_{\pi^+ p}=a_3$, $\tilde{a}_{cc}=(a_3+2 a_1)/3$, and $\tilde{a}_{c0}=\sqrt{2} (a_3-a_1)/3$. The $s$-wave part of the low-energy $\pi N$ scattering amplitude is of the form 
$b_0 + b_1 \vec{\tau} \cdot \vec{t}$, where $b_0$ and $b_1$ are the isoscalar and isovector scattering lengths, respectively, and $\vec{t}$ is the isospin operator of the pion. The quantity $b_0$ (frequently denoted as $a^+$ 
or $a_{0+}^+$) is related to $a_{cc}$ and $a_{c0}$ according to the formula $b_0=a_{cc}+a_{c0}/\sqrt{2}$, whereas $b_1$ (in several works in the domain of Pion Physics, $-b_1$ is denoted as $a^-$ or $a_{0+}^-$) is simply 
equal to $a_{c0}/\sqrt{2}$. After removing the EM contributions from $b_0$ and $b_1$, one obtains the isoscalar and isovector hadronic scattering lengths $\tilde{b}_0$ and $\tilde{b}_1$. The relations to the two scattering 
lengths in isospin basis read as: $\tilde{b}_0=(2 a_3+a_1)/3$ and $\tilde{b}_1=(a_3-a_1)/3$.

\section{\label{sec:Determinations}Determinations of the various $\pi N$ coupling constants}

The methods for determining the $\pi N$ coupling constant may be categorised on the basis of the theoretical model, which is used in order to describe the experimental data ($\pi N$, $N N$), and, of course, of the 
experimental input itself.
\begin{itemize}
\item Physical models of the $\pi N$ interaction and the $\pi N$ experimental data (differential and total/partial-total/total-nuclear cross sections, as well as analysing powers).
\item Physical (meson-exchange) models of the $N N$ interaction and the $N N$ (e.g., $p p$, $n p$, $\bar{p} p$) experimental data. Relevant in this case are Feynman graphs (simply graphs from now on) with exchanged pion(s) 
between the two interacting nucleons.
\item Dispersion-relation analyses, performed on the $\pi N$ and/or $N N$ experimental data.
\item Use of Current-Algebra constraints, of the GMO sum rule \cite{goldberger1955}, etc.
\end{itemize}

\subsection{\label{sec:OldDeterminations}Early determinations}

Although the first efforts to determine the $\pi N$ coupling constant took place in the beginning of the 1950s (see Section 2 of Ref.~\cite{swart1997}), the estimates were rather inaccurate for at least one 
decade~\footnote{Section 2 of Ref.~\cite{swart1997} provides an extensive list of the determinations of the $\pi N$ coupling constant before 1968; unfortunately, Ref.~\cite{hamilton1963} is not mentioned in that list.}. One 
of the very first accurate $f_c^2$ estimates (perhaps, the first one) appeared in Ref.~\cite{hamilton1963}. This interesting review paper also promulgates the use of forward dispersion relations for the $B$ amplitudes as 
``the most promising method for determining $f^2$'' (p.~762). From an analysis of $\pi N$ data, the authors obtained $f_c^2 =0.081(3)$.

Performing a dispersion-relation analysis of the then available $p p$ data, Bugg determined $f_p^2$ to $0.075(4)$ \cite{bugg1968} in 1968. Two years later, Ebel and collaborators \cite{ebel1970} placed $f_c^2$ between 
$0.076$ and $0.082$. In a subsequent paper, Brown and collaborators \cite{brown1971} found that their $f_c^2$ estimates, based on Current-Algebra constraints and the Adler-Weisberger theorem \cite{adler1965,weisberger1966}, 
ranged between $0.075$ and $0.080$. The authors favoured $f_c^2 = 0.077(2)$, where the uncertainty has been obtained by means of a comparison of their Eqs.~(25,34,37,41).

Applying fixed-$t$ dispersion relations to low-energy $\pi N$ differential (DCS) and total (TCS) cross sections, Bugg and collaborators \cite{bugg1973} obtained in 1973 an estimate for the $\pi N$ coupling constant, as well 
as estimates for the two scattering lengths for ES $\pi^\pm p \to \pi^\pm p$. Having been used (as input) in a variety of studies~\footnote{These studies are easily recognisable, as - if not directly quoting the $f_c^2$ 
result of Ref.~\cite{bugg1973} - they mention the use of $g^2/(4 \pi)=14.28$.}, the result $f_c^2=0.0790(10)$ has been one of the most influential in the domain of Hadronic Physics. For several decades, the value of 
Ref.~\cite{bugg1973} was acknowledged as `canonical', and (deplorably) still is for some. As de Swart and collaborators \cite{swart1997} remarked: ``We were surprised to note the many physicists trying to hold on to the 
old values.'' Be that as it may, one cannot but notice the very small $f_c^2$ uncertainty of Ref.~\cite{bugg1973}, which de Swart and collaborators \cite{swart1997} considered to be ``optimistic''; I cannot but endorse 
their opinion.

The results of the broadly-used partial-wave analysis (PWA) of Koch and Pietarinen \cite{koch1980} appeared in 1980. That solution became known as KH80 (the initials `KH' stand for `Karlsruhe' and `Helsinki'). In the 
abstract of their paper, the authors summarised their work: ``An energy-independent partial-wave analysis has been performed on pion-nucleon elastic and charge-exchange differential cross sections and elastic polarizations, 
for lab.~momenta below $500$ MeV/$c$ \dots For the pion-nucleon coupling constant the value $f_c^2=0.079(1)$ was obtained.''

Regarding the solution KH80, a number of remarks need to be made. To start with, very few low-energy data were available at the time when that analysis was performed. The trouble unfolds as one notices that, in the 
low-energy region, the analysis entirely relied on the $\pi^+ p$ DCSs of Bertin and collaborators \cite{bertin1976}. These seven data sets, each comprising ten measurements, have been criticised in all modern analyses of 
the $\pi N$ data; they prominently stand out from the bulk of the measurements~\footnote{There are three ways by which one could make use of the BERTIN76 data sets in a phase-shift analysis: a) by implementing a 
robust-analysis technique, b) by assigning a low weight to these data sets in optimisations featuring the conventional $\chi^2$ minimisation function, or c) by using these measurements in conjunction with a plethora of other 
data, enabling at the same time the rescaling (controlled floating) of the input data sets.}. One additional objection to the KH80 analysis relates to their omission of the normalisation uncertainties of the input data sets 
(see p.~336 of Ref.~\cite{koch1980}). Many researchers still make use of the solution KH80 without realising (or after turning a blind eye to) these shortcomings. Let me finally comment on the $f_c^2$ estimate of 
Ref.~\cite{koch1980}. Koch and Pietarinen provide some relevant details in Section 4.2 of their paper; as we will shortly see, not everyone agrees that these authors \emph{determined} $f_c^2$ in Ref.~\cite{koch1980}. I 
have few doubts that, though they attempted to `sell' this $f_c^2$ value as a determination in the abstract of their paper, they adroitly manoeuvred towards the reported value by letting themselves be steered by the result 
of Ref.~\cite{bugg1973}.

\subsection{\label{sec:After1980}Determinations between 1981 and 1997}

For about one decade, most researchers in the domain of Hadronic Physics believed that the consistency of the results between Refs.~\cite{bugg1973,koch1980} (which were taken for independent determinations) suggested that 
the question of the $\pi N$ coupling constant had been resolved and that attention could be diverted to other, more urgent matters, e.g., to the obvious discrepancies between the first modern (meson-factory) measurements 
of the $\pi^\pm p$ ES DCSs and the corresponding predictions obtained from the Karlsruhe analyses (i.e., KH80 and, performed by Koch in 1985, KA85). Fortunately, there were also those who had doubts, as (for instance) the 
case was with the Nijmegen group. Details on the development of the Nijmegen potentials, as well as a list of their estimates for the various $\pi N$ coupling constants over time, are given in Sections 3 and 4 of 
Ref.~\cite{swart1997}. I will now attempt to concisely reconstruct the Nijmegen story (all references may be found in Ref.~\cite{swart1997}). The description of the experimental data with the Nijmegen hard-core potential 
of 1975 resulted in $f^2 \approx 0.0741$. Their soft-core potential of 1978, along with constraints on the $f^2$ range of permissible values, yielded $f^2 \approx 0.0772$. (At this point, Ref.~\cite{swart1997} hints at the 
extraction of a smaller $f^2$ value, in case that an unconstrained fit had been performed - which, no doubt, would have been the authors' choice; the constrained fit had prevented the drop of the fitted $f^2$ value `beyond 
reason'.) After analysing $p p$ data, the Nijmegen group became gradually convinced that $f_p^2$ should be significantly smaller than the canonical value, and announced this supposition at the 1983 Few-Boby Conference in 
Karlsruhe. A few years later, an analysis of $p p$ data at $350$ MeV resulted in an accurate determination of $f_p^2$ to $0.0725(6)$, which was updated (around the end of the 1980s) to $0.0749(6)$. As the group had not yet 
extracted themselves an estimate for $f_c^2$, they relied on the use of the canonical value in their investigation of the violation of the isospin invariance (the preferred term for `isospin invariance' in the $N N$ sector 
is `charge independence'). Evidently, the comparison between their $f_p^2$ value and the canonical value yielded large isospin-breaking effects; the report of those effects was not received with enthusiasm.

It was in summer 1990 when Arndt and collaborators \cite{arndt1990} published an article favouring an $f_c^2$ value which was considerably smaller than the canonical value. After analysing the then available $\pi N$ ES 
measurements below $2$ GeV using fixed-$t$ dispersion relations (see also Ref.~\cite{arndt1991} for details on the solution which became known as SM90), the authors reported the result $f_c^2=0.0735(15)$ and commented 
further in the abstract of their paper: ``\dots a value in conflict with the result of Koch and Pietarinen, yet consistent with the value of the $\pi^0 p p$ coupling determined in the recent Nijmegen analysis of $p p$ 
scattering data.'' Although it had not really been ``the result of Koch and Pietarinen,'' the wheels had been set in motion.

The first accurate estimates by the Nijmegen group for all $\pi N$ coupling constants appeared in autumn 1991. Klomp and collaborators \cite{klomp1991} remarked in the abstract of that paper: ``The $N N \pi$ coupling 
constants are extracted in $N N$ partial-wave analyses. The data base contains all $p p$ and $n p$ scattering data below $T_{\rm lab}=350$ MeV. Introducing different coupling constants at the different $N N \pi$ vertices, 
at the pion pole we find for the $p p \pi^0$ coupling $f_p^2=0.0751(6)$, for the $n n \pi^0$ coupling $f_n^2=0.075(2)$, and for the charged-pion coupling $f_c^2=0.0741(5)$. These results allow only small 
charge-independence-breaking effects in the $N N \pi$ coupling constants. If we assume charge independence, we find $f^2=0.0749(4)$.'' To the best of my knowledge, that was the first statement by the Nijmegen group on the 
absence of significant splitting effects in the $\pi N$ coupling constant. The $f_0^2=0.0752(8)$ result of Ref.~\cite{klomp1991} was quoted in Ref.~\cite{swart1997}; I am not aware of a more recent $f_0^2$ result by the 
Nijmegen group. Additional details on the analysis may be found in Ref.~\cite{stoks1993}, an important paper featuring the precise result $f_c^2=0.0748(3)$, the final $f_c^2$ value by the Nijmegen group.

Using fixed-$t$ dispersion relations on $\pi N$ ES data for pion laboratory kinetic energy $T$ between $100$ and $310$ MeV, Markopoulou-Kalamara and Bugg \cite{markopoulou1993} obtained in 1993 a new accurate estimate for 
$f_c^2$: $0.0771(14)$, i.e., a value smaller than (yet not incompatible with) the 1973 result of Ref.~\cite{bugg1973}.

From a PWA of all $\bar{p} p$ scattering data below $925$ MeV/$c$ (antiproton laboratory momentum), Timmermans and collaborators \cite{timmermans1994} obtained $f_c^2=0.0732(11)$ in 1994; a subsequent analysis with an 
updated $\bar{p} p$ DB led to $f_c^2=0.0736(10)$ \cite{swart1997}. Also in 1994, Arndt and collaborators \cite{arndt1994a} performed PWAs of the $\pi N$ ES data up to $2$ GeV, using forward and fixed-$t$ dispersion 
relations, and obtained chi-square maps for fixed $g^2/(4 \pi)$ and $b_0$ values. Their preferred solution for $g^2/(4 \pi)$ was $13.75(15)$, translating into $f_c^2=0.0761(8)$. (It needs to be said that 
Ref.~\cite{arndt1994a}, known as solution FA93, uses another definition of $f^2$, not absorbing in it a factor $4 \pi$.) By the end of 1994, two groups of $f_c^2$ values had clearly been established: the pre-meson-factory 
group, comprising values obtained up to 1980 and centred around the canonical value, and the Nijmegen-VPI/GWU group, comprising values obtained in the early 1990s and centred around $0.075$. Those of us who portrayed this 
discrepancy as a `disagreement between the \emph{outdated} and the \emph{modern}' learnt in 1995 (the hard way) that the \emph{modern} is not necessarily self-consistent. Already in January, Bradamante and collaborators 
\cite{bradamante1995}, using $\bar{p} p \to \bar{n} n$ DCSs from the CERN Low Energy Antiproton Ring (LEAR), reported a very low estimate for $f_c^2$; the reported value was equal to $0.071(2)$. The year went on as 
promisingly as it had started. In August 1995, Ericson and collaborators \cite{ericson1995}, using $n p$ DCSs at $162$ MeV, acquired at the neutron beam facility at the Svedberg Laboratory in Uppsala, obtained 
$f_c^2=0.0808 \pm 0.0003 \pm 0.0017$, i.e., a large estimate for the charged-pion coupling constant, in support of the canonical value and in conflict with most of the values obtained from the $\pi N$ sector, as well as 
with the earlier result from LEAR.

About one month later, the paper of Bugg and Machleidt \cite{bugg1995} appeared, reporting the results of an analysis of high partial waves for $p p$ and $n p$ ES between $210$ and $800$ MeV. The authors remarked in the 
abstract of their paper: ``There are some discrepancies, but sufficient agreement that values of the $\pi N N$ coupling constants $g_0^2$ for $\pi^0$ exchange and $g_c^2$ for charged $\pi$ exchange can be derived. Results 
are $g_0^2=13.94 \pm 0.17 \pm 0.07$ ($p p$) and $g_c^2=13.69 \pm 0.15 \pm 0.24$ ($n p$), where the first error is statistical and the second is an estimate of the systematic error arising from uncertainties in the normalization 
of total cross sections and $d\sigma/d\Omega$.'' (In this paper, the factor $4 \pi$ has been absorbed in the quantities $g_0^2$ and $g_c^2$.) The two results translate into $f_c^2=0.0756(22)$ and $f_0^2=0.0771(13)$, where 
the statistical and systematic uncertainties of Ref.~\cite{bugg1995} have been linearly combined (i.e., summed). The $f_c^2$ estimate of Ref.~\cite{bugg1995} landed in-between the two earlier results of that year.

The $7^{\rm th}$ MENU (`Meson-Nucleon Physics and the Structure of the Nucleon') Conference took place in Vancouver in summer 1997. Before my contribution, de Swart gave an emotional talk on the status of the $\pi N$ 
coupling constant \cite{swart1997}, one of those talks which are bound to remain in one's memory; I will shortly comment further on that talk. Timmermans came afterwards \cite{timmermans1997}, reporting the result 
$f_c^2=0.0756(9)$, obtained from $\pi N$ data below $410$ MeV. I followed with the description of a robust analysis of the low-energy $\pi N$ measurements \cite{matsinos1997a,matsinos1997b} and reported: $f_c^2=0.0765(14)$. 
By the end of the conference, the general consensus of opinion was that the value of $\pi N$ coupling constant had to be significantly smaller than the canonical value, e.g., see the remarks in Ref.~\cite{wagner1997}.

I shortly return to de Swart's talk. The abstract of his paper with Rentmeester and Timmermans, which appeared in the proceedings of the conference, is indicative of the atmosphere which the talk itself created. The 
authors vividly state: ``A review is given of the various determinations of the different $\pi N N$ coupling constants in analyses of the low-energy $p p$, $n p$, $\bar{p} p$, and $\pi p$ scattering data. The most accurate 
determinations are in the energy-dependent partial-wave analyses of the $N N$ data. The recommended value is $f^2=0.075$. A recent determination of $f^2$ by the Uppsala group from backward $n p$ cross sections is shown to 
be model dependent and inaccurate, and therefore completely uninteresting \dots'' Regarding the KH80 $f_c^2$ determination, the authors clarify: ``The outstanding Karlsruhe-Helsinki partial-wave analyses of the $\pi N$ 
data used the value of Bugg \etal~as input. In 1980, Koch and Pietarinen \cite{koch1980} used fixed-$t$ dispersion relations and found again that $f_c^2=0.079(1)$. However, this is more a consistency check than a real 
determination, because the value of the coupling constant was used as \emph{input} in the analyses. Other values of $f_c^2$ were not tried as input.'' I must admit that de Swart's comments, as well as those few lines in 
Ref.~\cite{swart1997}, enjoined me to reread Ref.~\cite{koch1980}, with a more critical eye.

One could summarise the essential results of the Nijmegen programme in two sentences.
\begin{itemize}
\item No significant splitting effects have been observed in the (values of the) $\pi N$ coupling constant.
\item The recommended value for the $\pi N$ coupling constant lies in the vicinity of $0.075$, i.e., well below the canonical value.
\end{itemize}
According to Timmermans \cite{timmermans2018}, the results of Refs.~\cite{stoks1993,timmermans1997} for $f_c^2$ and the one of Ref.~\cite{rentmeester1999} for $f_p^2$ should be considered to be final by the Nijmegen group. 
The $f_c^2$ value of Ref.~\cite{timmermans1994} was slightly updated in Ref.~\cite{swart1997}. I will return to Ref.~\cite{rentmeester1999} shortly.

\subsection{\label{sec:RecentResults}Determinations after 1998}

Early in 1998, Gibbs and collaborators \cite{gibbs1998} obtained $f_c^2=0.0756(7)$ from an analysis of modern low-energy $\pi^\pm p$ ES data. In their determination, the authors made use of the GMO sum rule.

The most recent determination of $f_p^2$ by the Nijmegen group originates from 1999. Studying the long-range properties of the $p p$ interaction in an energy-dependent PWA, Rentmeester and collaborators \cite{rentmeester1999} 
obtained $f_p^2=0.0755(7)$.

The next two reports \cite{rahm1998,rahm2001} may be thought of as a follow-up (but autonomous) work of Ref.~\cite{ericson1995}. The former paper reports the results of an analysis of the $n p$ DCS at $162$ MeV between 
$72^\circ$ and $180^\circ$. The authors stress again that ``special attention was paid to the absolute normalization of the data.'' They also observe that ``in the angular range $150^\circ-180^\circ$, the data are steeper 
than those of most previous measurements and predictions from energy-dependent partial-wave analyses or nucleon-nucleon potentials. At $180^\circ$, the difference is of the order of $10-15$ \%.'' The authors finally report: 
$f_c^2=0.0803(14)$. Measurements of the $n p$ DCS at $96$ MeV (within almost the same angular range) were analysed in Ref.~\cite{rahm2001}, and the value $f_c^2=0.0814(18)$ was obtained. The results from Uppsala 
\cite{ericson1995,rahm1998,rahm2001} are consistent among themselves. Noticeable, of course, is the exclusive use of $n p$ data in all three works; in fact, these are the only works on the $\pi N$ coupling constant, which 
use only $n p$ DCSs as input.

The subsequent paper serves as the final report of the group which acquired the pioneering measurements of the strong shift $\epsilon_{1 s}$ and of the total decay width $\Gamma_{1 s}$ in pionic hydrogen and deuterium 
\cite{schroeder2001} at the Paul Scherrer Institute (PSI). In that paper, the EM effects in the case of pionic hydrogen were removed after using the results of an earlier work, which had been published by part of the group 
and Oades in 1996 \cite{sigg1996a}. Two quantities were imported from that work into Ref.~\cite{schroeder2001}, namely $\delta_\epsilon=-2.1(5) \cdot 10^{-2}$ (relating to the correction which one needs to apply to 
$\epsilon_{1 s}$) and $\delta_\Gamma=-1.3(5) \cdot 10^{-2}$ (relating to the removal of the EM effects from $\Gamma_{1 s}$).

The trouble with the results of Ref.~\cite{sigg1996a} is that they stemmed from a two-channel calculation ($\pi^- p \to \pi^- p, \pi^0 n$), along with the \emph{phenomenological} addition of the contributions of the `third' 
channel ($\pi^- p \to \gamma n$). The three-channel calculation \cite{oades2007}, performed a few years after Ref.~\cite{schroeder2001} appeared, resulted in a significantly different result for $\delta_\epsilon$. To be fair, 
I remember that, on a number of occasions even before Ref.~\cite{sigg1996a} was published, Oades had voiced his reservations about the contributions of the $\gamma n$ channel in the case of $\epsilon_{1 s}$. In Appendix 
\ref{App:AppA}, I present results obtained after the application of different sets of corrections to the experimental results for $\epsilon_{1 s}$ and $\Gamma_{1 s}$ of Ref.~\cite{schroeder2001}, after updating the former 
result by using a more recent theoretical determination of the $3 p \to 1 s$ EM transition energy.

In Ref.~\cite{schroeder2001}, the authors obtained an estimate for the $\pi N$ coupling constant (namely, $g_{\pi N}=13.21^{+0.11}_{-0.05}$) from the isovector hadronic scattering length $\tilde{b}_1$ by use of the GMO sum 
rule. To obtain the $\tilde{b}_1$ estimate, Schr{\"o}der and collaborators combined information from pionic hydrogen and deuterium. In view of the fact that the EM corrections in the former case \cite{sigg1996a} appear 
incomplete, I will refrain from including the authors' $f^2$ value in the list of results which are analysed in Section \ref{sec:DeterminationsOfAverages}.

In 2002 and 2004, Ericson and collaborators published two papers \cite{ericson2002,ericson2004} using results from Ref.~\cite{schroeder2001} as input. The former paper imported the result of Ref.~\cite{schroeder2001} for 
the scattering length $\tilde{a}_{cc}$. In the second paper, the authors turned a critical eye on the EM corrections of Ref.~\cite{sigg1996a}. They comment that the potential-model approach is ``model dependent'' and, 
furthermore, they find it inconsistent with their low-energy expansion \cite{ericson2002}. The authors derived the EM corrections at threshold on the basis of their own model of the $\pi^- p$ atom; numerical results may be 
found in their Table 1. Their $\delta_\epsilon$ value lies in-between the results of Refs.~\cite{sigg1996a,oades2007}, slightly closer to the latter result, whereas their $\delta_\Gamma$ value is of opposite sign and 
disagrees with both works \cite{sigg1996a,oades2007}. The result of Ref.~\cite{ericson2004} for $g_c^2 / (4 \pi)$ was $14.04(17)$, which translates into $f_c^2=0.07756(94)$. This value will be included in the analysis 
presented in Section \ref{sec:DeterminationsOfChargedCoupling}.

Starting from their solution FA93, the $f_c^2$ results of the VPI/GWU group have been remarkably stable over the years. The sensitivity of the results of their various analyses to $f_c^2$ was investigated in several papers 
\cite{arndt1994b,arndt2000,pavan2000} after their solution FA93 appeared. Figure 5 of their paper \cite{arndt1994b} raises the possibility that the $f_c^2$ results, obtained from separate analyses of $p p$ and $n p$ data, 
might not match well and lie on either side of the $\chi^2$ minimum corresponding to the $\pi N$ data. Although the authors mention on p.~2737 that this picture depends on the details of the one-pion exchange mechanism, it 
cannot be excluded that this observation might provide an explanation for the fluctuation in the $f_c^2$ values of Fig.~\ref{fig:ChargedCouplingConstant} of this work. In addition, Fig.~5 of Ref.~\cite{arndt1994b} 
demonstrates that the optimisation to the \emph{combined} $p p$ and $n p$ DBs yields results compatible with those obtained from the $\pi N$ data, though the sensitivity of the analysis to $f_c^2$ is still low. Reference 
\cite{pavan2000} gives the result $g_c^2 / (4 \pi)=13.73 \pm 0.01 ({\rm stat.}) \pm 0.07 ({\rm syst.})$ and remarks that this value was ``found to be insensitive to database changes and Coulomb barrier corrections.'' The 
subsequent solution by the VPI/GWU group \cite{arndt2004}, i.e., their solution FA02, was obtained from a simultaneous fit of $\pi^\pm p$ ES and CX data up to $T=2.1$ GeV, as well as of $\pi^- p \to \eta n$ data up to $0.8$ 
GeV. Therein, the $g_c^2 / (4 \pi)$ value of $13.75(10)$ was obtained, i.e., the central value of the solution FA93, accompanied by a sizeably smaller uncertainty; I will consider this value, namely $f_c^2=0.07596(55)$, to 
be the final result of the VPI/GWU group regarding the context of this work. Their subsequent solution SP06 \cite{ardnt2006} yielded a perfectly compatible result for the central $f_c^2$ value.

Performing a re-analysis of $\pi N$ data and dispersion relations for the isoscalar invariant amplitude $B^+$, Bugg obtained in 2004 a new estimate for $g_c^2 / (4 \pi)$ \cite{bugg2004}, which translates to $f_c^2 = 0.07590(55)$.

In their 2011 paper, Baru and collaborators \cite{baru2011a} obtained $g_c^2 / (4 \pi)=13.69(20)$ from PSI results on pionic hydrogen and deuterium, using the GMO sum rule. As I have major objections to the authors' choice 
of input, I will refrain from including their $g_c^2 / (4 \pi)$ estimate in the analysis of Section \ref{sec:DeterminationsOfAverages}. To start with, the authors chose to use in their study \emph{preliminary} $\epsilon_{1 s}$ 
and $\Gamma_{1 s}$ results from a PSI experiment, which was the follow-up experiment of the mid-1990s experiment by the ETHZ-Neuch{\^a}tel-PSI Collaboration, rather than the \emph{final} results of that first experiment 
\cite{schroeder2001}. It so happened that the final $\epsilon_{1 s}$ result of the follow-up experiment by the Pionic-Hydrogen Collaboration (which is beautifully compatible with the result of Ref.~\cite{schroeder2001}, but 
considerably more accurate) appeared a few years later \cite{hennebach2014} and was somewhat smaller (in absolute value) than the value which had been used as input in Ref.~\cite{baru2011a}. (Reference \cite{hennebach2014} 
uses another sign convention for $\epsilon_{1 s}$.) My major objection to Ref.~\cite{baru2011a} relates to $\Gamma_{1 s}$; to the best of my knowledge, the Pionic-Hydrogen Collaboration have not yet published a final result! 
In their 2015 paper \cite{gotta2015}, their estimate for $\Gamma_{1 s}$ of pionic hydrogen~\footnote{It is frequently assumed that the pionic-hydrogen data suffer from fewer problems in comparison with the measurements of 
the DCS above threshold; this assumption might be fallacious. The determination of $\Gamma_{1s}$ in pionic hydrogen, as emerging from the experimental activity of the Pionic-Hydrogen Collaboration at PSI, is a manifestation 
of such problems. Preliminary results for $\Gamma_{1s}$ were announced at a conference in 2015 \cite{gotta2015}, over one decade after the experiments (on pionic hydrogen) were completed. However, no concrete picture emerges 
from Fig.~3 of that paper for $\Gamma_{1s}$: the corresponding probability distributions, obtained from three transitions ($2 p, 3 p, 4 p \to 1 s$), each measured twice, hardly overlap. In a perfect world, and assuming that 
all effects have been taken into account correctly, these six distributions should agree, as they all represent the total decay width of the ground state of pionic hydrogen.} was still marked as `preliminary' and was given 
as $\Gamma_{1 s}=850^{+40}_{-50}$ meV, i.e., slightly more accurate than the $\Gamma_{1 s}$ result of Ref.~\cite{schroeder2001}. Baru and collaborators \cite{baru2011a} had used $\Gamma_{1 s} = 0.823(19)$ eV in their 2011 
paper, i.e., a smaller value, accompanied by a largely underestimated uncertainty.

Within an $N N$ model based on one-pion exchange, Babenko and Petrov \cite{babenko2016} reported $g_c^2 / (4 \pi) = 14.55(13)$ and $g_0^2 / (4 \pi) = 13.55(13)$. Their two results suggest the violation of the isospin 
invariance, which (as we saw earlier) the Nijmegen analyses refute. As the authors remark, their $g_c^2 / (4 \pi)$ value is consistent with those reported by the Uppsala group \cite{ericson1995,rahm1998,rahm2001}. To the 
best of my knowledge, the works of Babenko and Petrov are the only ones which suggest that $f_c>f_0$. The authors slightly updated their results one year later to $g_c^2 / (4 \pi) = 14.53(25)$ and $g_0^2 / (4 \pi) = 13.52(23)$ 
\cite{babenko2017}; the updated values (i.e., $f_c^2=0.0803(14)$ and $f_0^2=0.0748(13)$) will be used here.

In their 2016 paper, Ruiz Arriola and collaborators \cite{ruiz2016} discussed their results $f_p^2=0.0759(4)$, $f_0^2=0.079(1)$, and $f_c^2=0.0763(6)$, extracted earlier from a PWA of a DB which the authors call ``$3 \sigma$ 
self-consistent $N N$ database'' (Granada-2013) comprising $6713$ measurements acquired between 1950 and 2013. The results were slightly updated in 2017 \cite{navarro2017}, where the authors report the values $f_p^2=0.0761(4)$, 
$f_0^2=0.0790(9)$, and $f_c^2=0.0772(6)$, obtained from almost the same data as their earlier work; the updated values will be used here. Interestingly, the authors draw attention to the (large) anticorrelation between 
$f_c^2$ and $f_0^2$ in their analysis. The authors' estimates for $f_c^2$ and $f_0^2$ do not match the results of Ref.~\cite{babenko2017}.

\section{\label{sec:DeterminationsOfAverages}Determinations of averages for the various $\pi N$ coupling constants}

\subsection{\label{sec:DeterminationsOfChargedCoupling}Determinations of $f_c^2$}

The $\pi N$ coupling constant $f_c^2$ is the one with the most determinations (or, better expressed, attempts at a determination); in total, twenty reported values fulfil the criteria put forward in Section 
\ref{sec:Introduction}. Analysed in this section are the ten $f_c^2$ values of Refs.~\cite{ebel1970,stoks1993,timmermans1994,bradamante1995,ericson1995,bugg1995,rahm1998,rahm2001,babenko2017,navarro2017} from the $N N$ 
system and the ten values of Refs.~\cite{hamilton1963,brown1971,bugg1973,markopoulou1993,timmermans1997,matsinos1997b,gibbs1998,ericson2004,arndt2004,bugg2004} from the $\pi N$ system. As mentioned in Section \ref{sec:After1980}, 
the value of Ref.~\cite{timmermans1994} was updated in Ref.~\cite{swart1997}; although the difference between the two values is small, the updated result will be used.

\begin{figure}
\begin{center}
\includegraphics [width=15.5cm] {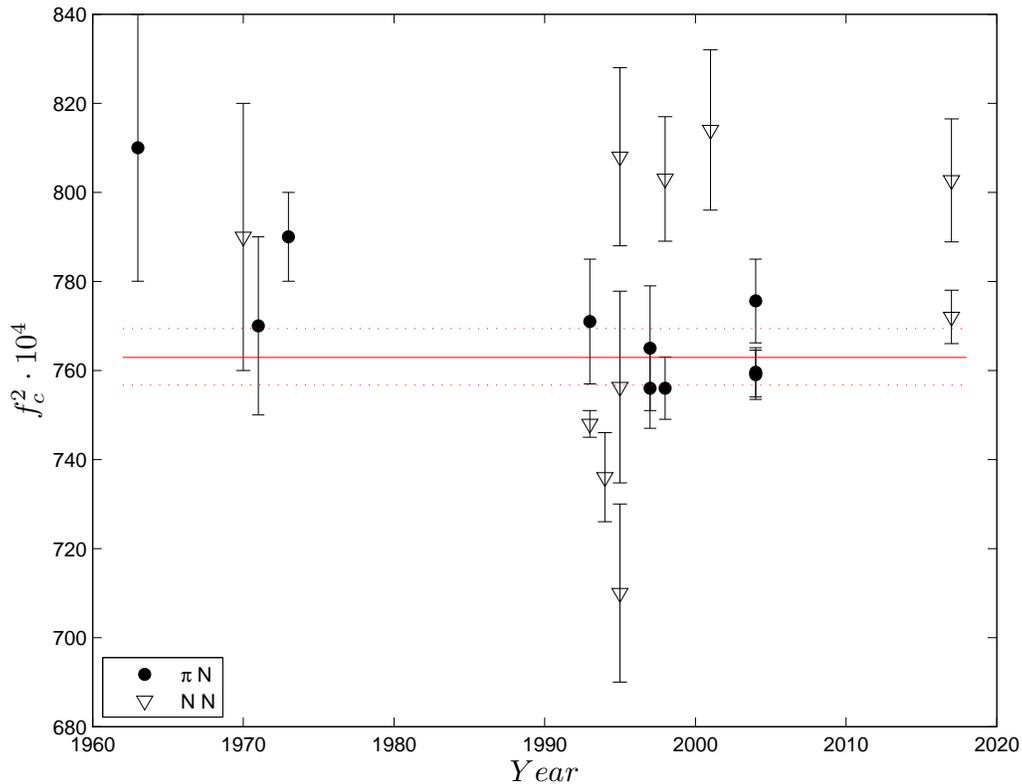}
\caption{\label{fig:ChargedCouplingConstant}The results for the $\pi N$ coupling constant $f_c^2$ from $\pi N$ and from $N N$ analyses, separately shown. The solid straight line represents an average of the displayed values, 
obtained from robust fits. The two dotted straight lines represent the (asymmetrical) $1 \sigma$ uncertainties around that average, see Eq.~(\ref{eq:EQ03}).}
\vspace{0.35cm}
\end{center}
\end{figure}

Figure \ref{fig:ChargedCouplingConstant} contains all the results, separately for the estimates originating from the $\pi N$ and from the $N N$ analyses. Poring over this figure without knowing what is being plotted, one 
might doubt that the data points correspond to estimates for the same physical quantity. On the other hand, the trouble with the fluctuation in the figure lies with the $f_c^2$ estimates obtained from the $N N$ data; those 
extracted from the $\pi N$ data cluster around their weighted average in an acceptable way~\footnote{Using only the $f_c^2$ estimates from the $\pi N$ data, one obtains from a simple $\chi^2$ (one-parameter) fit: 
$f_c^2=0.07635(35)$; the resulting $\chi^2$ of this fit is about $14.49$ for $9$ degrees of freedom (DoF), corresponding to a p-value of $1.07 \cdot 10^{-1}$, i.e., well above the threshold $\mathrm{p}_{\rm min} = 1.00 \cdot 10^{-2}$, 
regarded by most statisticians as the outset of statistical significance.}. Although it has been suggested in the past that the best determinations of $f_c^2$ should come from the $N N$ sector, Fig.~\ref{fig:ChargedCouplingConstant} 
can hardly substantiate such a claim.

For the analyst, Fig.~\ref{fig:ChargedCouplingConstant} is a nightmare. I had been considering for a while which procedure I could implement in order to obtain a meaningful average from such a spread of values (and of 
associated uncertainties). Because of the one-sided outliers (large-$f_c^2$ data points), it seemed to me reasonable to apply a robust technique.

One category of robust optimisations rest upon the use of the conventional $\chi^2$ minimisation function and the application of hard or soft weights to the input data points. At each (iteration) step in the optimisation 
scheme, the software application, which drives the function minimisation, varies the fit parameters (according to dedicated algorithms) and passes each new vector of parameter values to the user-defined function which hosts 
the parametric (theoretical) model. Fitted values (corresponding to the input vector of parameter values at that step) are generated within this function for all input data points. The distance between the input and the 
fitted values is evaluated for each data point. Hard-weight techniques use this distance to decide whether an input data point is an ordinary one or an outlier (at that step). Such an optimisation scheme is dynamical, in 
that data points which are outliers at one step may become ordinary at the next; similarly, ordinary points may turn into outliers from one step to another. These interchanges are more frequent in the initial phases of the 
optimisation, when the changes of the parameter values are larger. The nub of the matter is that the distances between the input and the fitted values (or, as the case is with measurements in Physics, the distances divided 
by the input uncertainties, i.e., the normalised residuals $r_i$) determine whether each data point is an ordinary point or an outlier, and also fix the weights of the $\chi^2$ contributions of all input data points at all 
steps.

Both hard- and soft-weight techniques evaluate a weight for each input data point at each step of the optimisation. Their difference is that the $\chi^2$ contributions from the outliers are turned into $0$ in the former 
case; the outliers are excluded. On the other hand, soft-weight techniques apply non-zero weights to the outliers, allowing them to participate at all steps of the optimisation. Hard and soft weights may be continuous or 
discontinuous. A typical example of a discontinuous hard-weight scheme would be to apply the weight of $0$ to all outliers and that of $1$ to all ordinary points. An example of a continuous hard-weight scheme would emerge 
if the input data points are categorised as follows: of type (a) are the ordinary points which lie within a distance $\epsilon_1>0$ of the fitted values, i.e., those satisfying $\lvert r_i \rvert \leq \epsilon_1$; of type 
(c) are the outliers, characterised by $\lvert r_i \rvert \geq \epsilon_2$, where $\epsilon_2>\epsilon_1$; and of type (b) are the ordinary points which satisfy $\epsilon_1 < \lvert r_i \rvert < \epsilon_2$, neither `too 
good' nor outliers. Weights of $1$ could be assigned to type (a), $0$ to type (c), and between $0$ and $1$ to type (b). The weights $W_i$ in the case of the type-(b) points may be chosen in such a way as to be continuous, 
monotonic, and fulfilling
\begin{equation*}
\lim_{\lvert r_i \rvert \to \epsilon_1^+} W_i = 1
\end{equation*}
and 
\begin{equation*}
\lim_{\lvert r_i \rvert \to \epsilon_2^-} W_i = 0 \, \, \, .
\end{equation*}
It is simple to pass into soft-weight schemes by devising a scheme which assigns a small (non-zero), monotonic weight to the outliers.

The approach of the present work relies on the use of seven continuous soft-weight robust optimisations of the reproduction of the data, displayed in Fig.~\ref{fig:ChargedCouplingConstant}, by one constant. Each of these 
methods contains one parameter, the so-called tuning parameter $k$. Although it is an adjustable scale, Statistics provides default values for $k$ for each method separately (obtained on the assumption that the residuals 
$r_i$ are normally distributed); these default values will be used. For the purpose of the function minimisation, the MINUIT package \cite{james} of the CERN library (FORTRAN version) was used. The weights, applied to the 
data points, purport to decrease the contributions of the data points which yield large $\lvert r_i \rvert$ contributions; all these contributions will be significantly smaller that they would have been, had a conventional 
$\chi^2$ minimisation function been used. For the sake of convenience, a new variable will be introduced, involving the default value of the tuning parameter $k$ of each method: $z_i=r_i/k$; the weights may thus be thought 
of as functions of $z_i$, rather than of $r_i$. In all cases, the weight in these optimisations (detailed below in alphabetical order) is set to $1$ for vanishing $z_i$. For non-zero values of the residuals ($z_i \neq 0$), 
the weights are set as follows:
\begin{itemize}
\item Andrews (constant contribution to the minimisation function for large $\lvert r_i \rvert$); default value of the tuning parameter $k=1.339$
\begin{equation} \label{eq:EQ02_01}
W_i (z_i) = \left\{
\begin{array}{rl}
2 z_i^{-2} \left( 1 - \cos \left( z_i \right) \right) & \text{, if $\lvert z_i \rvert < \pi$}\\
4 z_i^{-2} & \text{, otherwise}\\
\end{array} \right.
\end{equation}
\item Cauchy; default value of the tuning parameter $k=2.385$
\begin{equation} \label{eq:EQ02_02}
W_i (z_i) = z_i^{-2}\ln \left( 1 + z_i^2 \right)
\end{equation}
\item Fair; default value of the tuning parameter $k=1.400$
\begin{equation} \label{eq:EQ02_03}
W_i (z_i) = 2 z_i^{-2}\left( \lvert z_i \rvert - \ln \left( 1 + \lvert z_i \rvert \right) \right)
\end{equation}
\item Huber; default value of the tuning parameter $k=1.345$
\begin{equation} \label{eq:EQ02_04}
W_i (z_i) = \left\{
\begin{array}{rl}
1 & \text{, if $\lvert z_i \rvert < 1$}\\
z_i^{-2} \left( 2 \lvert z_i \rvert - 1 \right) & \text{, otherwise}\\
\end{array} \right.
\end{equation}
\item Logistic; default value of the tuning parameter $k=1.205$
\begin{equation} \label{eq:EQ02_05}
W_i (z_i) = 2 z_i^{-2} \ln \left( \cosh \left( z_i \right) \right)
\end{equation}
\item Tukey (constant contribution to the minimisation function for large $\lvert r_i \rvert$); default value of the tuning parameter $k=4.685$
\begin{equation} \label{eq:EQ02_06}
W_i (z_i) = \left\{
\begin{array}{rl}
(3 z_i^2)^{-1} \left( 1 - \left( 1 - z_i^2 \right)^3 \right) & \text{, if $\lvert z_i \rvert < 1$}\\
(3 z_i^2)^{-1} & \text{, otherwise}\\
\end{array} \right.
\end{equation}
\item Welsch; default value of the tuning parameter $k=2.985$
\begin{equation} \label{eq:EQ02_07}
W_i (z_i) = z_i^{-2} \left( 1 - \exp(- z_i^2) \right)
\end{equation}
\end{itemize}
In all cases, the aforementioned weight functions, which are continuous $\forall z_i \in \mathbf{R}$, guarantee that the corresponding seven minimisation functions follow the conventional $\chi^2$ function for small 
$\lvert r_i \rvert$ values. On the other hand, compared to the conventional $\chi^2$ function, the relevant contributions are reduced for large $\lvert r_i \rvert$. For the sake of example, the contribution to the 
conventional $\chi^2$ function is equal to $36$ when $\lvert r_i \rvert = 6$, whereas the contributions range between $7.17$ (Andrews) and $14.33$ (Huber) when using the weights detailed in Eqs.~(\ref{eq:EQ02_01}-\ref{eq:EQ02_07}).

\begin{figure}
\begin{center}
\includegraphics [width=15.5cm] {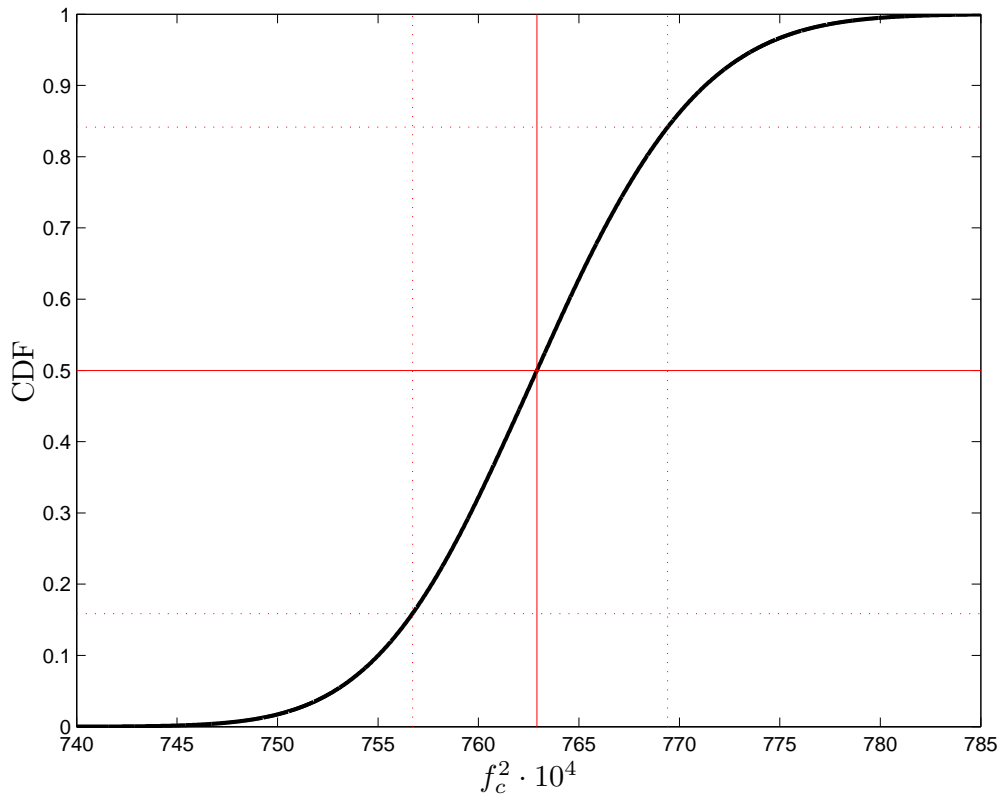}
\caption{\label{fig:CumulativeDistribution}The cumulative distribution function of $f_c^2$, emerging from a Monte-Carlo generation of random events in normal distribution (with asymmetrical uncertainties), utilising the 
results of the optimisation using the weights for the normalised residuals detailed in Eqs.~(\ref{eq:EQ02_01}-\ref{eq:EQ02_07}). The horizontal solid straight line corresponds to the $0.5$ level, yielding the median of 
the distribution (vertical solid straight line). The two horizontal dotted straight lines correspond to levels of about $0.1587$ and $0.8413$, thus yielding the $1 \sigma$ uncertainties around the median value (confidence 
level $\approx 68.27$ \%).}
\vspace{0.35cm}
\end{center}
\end{figure}

The results of the optimisation, using the aforementioned seven methods, are compatible within the fitted uncertainties. The asymmetrical uncertainties were obtained with the MINUIT method MINOS and were corrected for the 
quality of each fit via the application of the Birge factor (called scale factor by the Particle-Data Group). They were then used as input in the determination of the cumulative distribution function of $f_c^2$, obtained 
via the generation of $2.1$ billion Monte-Carlo events (i.e., $300$ million per optimisation method) and displayed in Fig.~\ref{fig:CumulativeDistribution}.

My recommendation as a representative average of the $f_c^2$ estimates of Fig.~\ref{fig:ChargedCouplingConstant} is:
\begin{equation} \label{eq:EQ03}
f_c^2 = 762.9^{+6.5}_{-6.2} \cdot 10^{-4} \, \, \, .
\end{equation}

Other efforts to extract a meaningful $\avg{f_c^2}$ value from the data of Fig.~\ref{fig:ChargedCouplingConstant} were carried out. A linear (one-parameter) least-squares fit, weighted only with the uncertainties of the 
input $f_c^2$ values (i.e., using $W_i=1$), yielded $\avg{f_c^2}=0.07595$, which is compatible with the result of Eq.~(\ref{eq:EQ03}); smaller fitted uncertainties than those quoted in Eq.~(\ref{eq:EQ03}) were obtained in 
this fit. Evidently, the estimate for $\avg{f_c^2}$ from this fit is pulled `downwards' by the $f_c^2=0.0748(3)$ result of Ref.~\cite{stoks1993}, which is accompanied by the smallest input uncertainty.

The fluctuation of the $f_c^2$ values in Fig.~\ref{fig:ChargedCouplingConstant} is sizeable. To be able to pinpoint the cause of this wide spread, a short, general outline of the procedure for the extraction of the various 
$\pi N$ coupling constants from the experimental data would be helpful.
\begin{itemize}
\item Experimental values of the standard observables (DCSs, TCSs, analysing powers, etc.), corrected for effects relating to beam contamination, target composition, detector efficiency, etc.~comprise the input into the 
analyses. Of crucial importance is the absolute normalisation of each input data set. The responsibility for the application of these corrections lies with the experimenters who acquired the measurements.
\item Removal of the EM effects, to extract hadronic quantities from the measurements. This is predominantly a task for theorists.
\item Modelling of the hadronic interaction, analysis technique. Relevant at this point is the model dependence of the results. I do not believe that the fluctuation can be accounted for by glitches in planning and applying 
the analysis technique.
\end{itemize}

If I were asked to single out one of the aforementioned possibilities as the most probable source of the fluctuation observed in Fig.~\ref{fig:ChargedCouplingConstant}, I would rather opt for the model dependence of the 
results. Two studies, using hadronic models with the same physical content and consistent input data, are bound to produce compatible results.

My second guess for the fluctuation in Fig.~\ref{fig:ChargedCouplingConstant} would be the model dependence of the removal of the EM effects from the measurements (or from the scattering amplitudes obtained thereof). I am 
not aware of comparative studies addressing the differences among the various schemes of application of the EM corrections both in the $\pi N$ and in the $N N$ sectors~\footnote{Regarding the $\pi N$ sector, there are as 
many such schemes as research programmes and, even worse, information about how the EM effects are treated in these schemes is either sparse or non-existing. Visual inspection of Table \ref{tab:EMCorrections} attests to 
the lack of a consensus on the EM corrections which one needs to apply even to the measurements obtained at the $\pi N$ threshold. Corrections which should (in principle) be compatible (as those discussed in Sections 
\ref{sec:PotentialModels} and \ref{sec:ELWModel}) disagree and even differ in sign. The corrections obtained within the framework of Chiral Perturbation Theory for the strong shift of the $1 s$ state in pionic hydrogen are 
large (when compared to the experimental uncertainties, as well as to the magnitude - absolute value - of the effects obtained in Sections \ref{sec:PotentialModels} and \ref{sec:ELWModel}) and, because of the poorly known 
low-energy constant $f_1$, very uncertain. To the best of my knowledge, only the Aarhus-Canberra-Zurich Collaboration has attempted the determination of the EM corrections for the scattering data (above threshold) and at 
threshold in a consistent manner (i.e., using the same potentials). I strongly believe that the problem of the EM corrections in $\pi N$ scattering must be revisited; the data analysis necessitates the availability of a 
consistent and reliable set of EM corrections from threshold up to the energy of a few GeV. The outstanding work of the NORDITA group \cite{tromborg1976,tromborg1977,tromborg1978} during the 1970s must be upgraded, after 
taking into account both the theoretical advancements, as well as the entirety of the experimental information which became available from the meson factories after 1980.}.

My third guess for the fluctuation in Fig.~\ref{fig:ChargedCouplingConstant} would be the (generally) underestimated systematic uncertainties associated with the experimental values, i.e., the normalisation uncertainties 
of each data set. The normalisation uncertainties of the data sets are closely linked to the uncertainties of the outcome of an analysis of the measurements. My experience suggests to me that, when providing estimates for 
the normalisation uncertainty in their experiments, most experimental groups tend to be on the optimistic side. One good example may be taken from the analyses of the low-energy $\pi N$ DBs. Scatter plots of the reported 
normalisation uncertainty (per experimental group and type of experiment) and the time (when the experiment was conducted) are generally expected to have a negative slope: on average, the experimental group is expected to 
gain experience with time, perfect their techniques, hence have a better grasp on the absolute normalisation of their data sets. In fact, the opposite tendency is observed on some occasions. This observation suggests to me 
that new sources of uncertainty surfaced with evolving time, evidently indicating that the normalisation uncertainties of the earlier works of that experimental group had been underestimated. As the experimenters bear the 
responsibility for updating their results (which, compared to the past, is straightforward and efficient nowadays), the only action, left to the analyst, is to wait for that moment to come.

To summarise, the fluctuation in Fig.~\ref{fig:ChargedCouplingConstant} might be explained on the basis of any of the following reasons (or their combination):
\begin{itemize}
\item[a)] Significant differences in the modelling of the hadronic part of the interaction.
\item[b)] Significant differences in the scheme of removal of the EM corrections.
\item[c)] Inconsistencies in the input measurements: for instance, erroneous determination of the absolute normalisation or underestimated normalisation uncertainties of the data sets.
\end{itemize}
Let me next use four selected $N N$ analyses and discuss how the differences in their estimates might be explained in the light of the three aforementioned possibilities.
\begin{itemize}
\item The Nijmegen $N N$ analyses were based on a large number of input data points and involved a large number of model parameters: their $1951$ $p p$ measurements below $350$ MeV were described by $22$ parameters 
\cite{rentmeester1999}, their $3964$ $n p$ measurements below $500$ MeV by $38$ parameters, and their $3847$ $\bar{p} p$ measurements by $36$ parameters \cite{swart1997}. In all cases, reduced-$\chi^2$ values close to $1$ 
have been reported.
\item The results from Uppsala \cite{ericson1995,rahm1998,rahm2001} were based on a small DB, comprising medium- and large-angle $n p$ DCSs at two energies: the three analyses were based on three data sets containing $31$, 
$54$, and $53$ data points, respectively.
\item Reference \cite{babenko2017} was based on low-energy $p p$ and $n p$ data. I could not retrieve from Refs.~\cite{babenko2016,babenko2017} details on the input, but the quoted fitted uncertainties indicate that the 
authors had used an extensive DB.
\item Reference \cite{navarro2017} is very straightforward in relation to their input data. Already in the abstract of their paper, one reads that their results were obtained ``from the Granada-2013 $n p$ and $p p$ database 
comprising a total of $6720$ scattering data below laboratory energy of $350$ MeV.'' Their Tables IV and V detail the $\chi^2$ contributions for the different observables; their reduced-$\chi^2$ values are between reasonable 
and excellent.
\end{itemize}
It goes without saying that the $f^2$ determinations in all four cases are subject to the effects of cases (a) and (b) above: different models and EM-correction schemes have been used in order to extract the $f^2$ estimates. 
In all probability, any inconsistencies in the input measurements would equally affect the Nijmegen analyses and those of Refs.~\cite{babenko2017,navarro2017}, hence the differences in their reported $f^2$ estimates cannot 
be traced to the effects of type (c) above. In any case, unless systematic, absolute-normalisation effects cancel out for input DBs comprising a large number of experimental data sets; the statistical expectation is that, 
for half of the input data sets, the absolute normalisation had been underestimated, whereas, for the other half, it had been overestimated. On the contrary, having been based on only three data sets, the results of 
Refs.~\cite{ericson1995,rahm1998,rahm2001} are much more sensitive to the effects of type (c) above.

\subsection{\label{sec:DeterminationsOfNeutralCoupling}Determinations of $f_0^2$}

\begin{figure}
\begin{center}
\includegraphics [width=15.5cm] {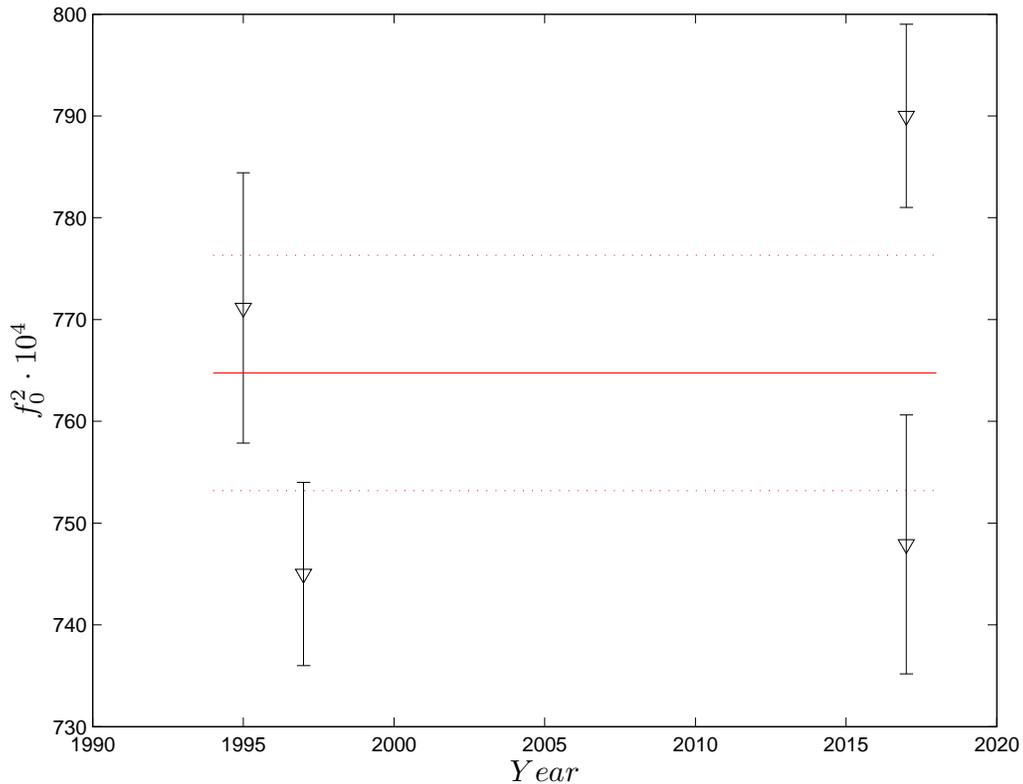}
\caption{\label{fig:NeutralCouplingConstant}The results for the $\pi N$ coupling constant $f_0^2$. The solid straight line represents the weighted average $f_0^2=0.0765(12)$ of the values displayed. The two dotted straight 
lines represent $1 \sigma$ uncertainties around the average.}
\vspace{0.35cm}
\end{center}
\end{figure}

There are only four determinations of $f_0^2$, all from $N N$ measurements, see Refs.~\cite{bugg1995,swart1997,babenko2017,navarro2017}. These values are displayed in Fig.~\ref{fig:NeutralCouplingConstant}. The weighted 
average comes out as
\begin{equation} \label{eq:EQ04}
f_0^2=0.0765(12) \, \, \, ,
\end{equation}
but the $\chi^2$ of the reproduction of these data (by one constant) is poor: $\chi^2 \approx 14.67$ for $3$ DoF (corresponding to a p-value of about $2.12 \cdot 10^{-3} < \mathrm{p}_{\rm min}$).

It is also interesting to investigate the difference $\Delta f^2 \coloneqq f_0^2-f_c^2$, as emerging from the analyses which reported both coupling constants. In the works of Refs.~\cite{bugg1995,swart1997}, there is no 
indication that $\Delta f^2 \neq 0$: in the former case, $\Delta f^2=15(25) \cdot 10^{-4}$; in the latter, $\Delta f^2=(-3.0 \pm 9.5) \cdot 10^{-4}$. On the other hand, an effect at the level of $2.9 \sigma$ is observed in 
the values of Ref.~\cite{babenko2017}, $\Delta f^2=-55(19) \cdot 10^{-4}$, suggesting that $f_0<f_c$. However, Ref.~\cite{navarro2017} does not support this finding; their value $\Delta f^2=18(11) \cdot 10^{-4}$ slightly 
favours $f_0>f_c$.

\subsection{\label{sec:DeterminationsOfOtherCouplings}Determinations of $f_p^2$ and $f_n^2$}

The $f_p^2$ determinations of Refs.~\cite{bugg1968,rentmeester1999,navarro2017} are well compatible. An average of these estimates (if an average of only three values makes any sense) is $f_p^2 = 0.07595(35)$, in agreement 
with the weighted average of $f_c^2$ in Eq.~(\ref{eq:EQ03}). Only one $f_n^2$ value is available~\footnote{Of course, this applies to the reported results which have directly been obtained from fits to measurements. 
Estimates for $f_n$ may be extracted indirectly, from $f_0$ and $f_p$ in the studies where both coupling constants were determined.}, that of Ref.~\cite{klomp1991}; its large uncertainty makes it compatible with all 
aforementioned averages.

\section{\label{sec:ETH}Determinations of the $\pi N$ coupling constant with the ETH model}

The ETH model of the $\pi N$ interaction (see Ref.~\cite{matsinos2014} and the references therein) is an isospin-invariant~\footnote{In the graphs of the model, the nucleon is assigned the proton mass, whereas the hadronic 
mass of the pion is taken to be the charged-pion mass. External mass differences are taken care of by the EM corrections applied to the $\pi N$ scattering amplitudes, as well as to the $\pi N$ phase shifts. Regarding 
Eq.~(\ref{eq:EQ02}), the ETH model uses $m_1=m_2=m_p$.} hadronic model based on $\sigma$- and $\rho$-meson $t$-channel exchanges, as well as on the $s$- and $u$-channel graphs with $N$ and $\Delta(1232)$ intermediate states. 
The contributions of all well-established $s$ and $p$ higher resonances with masses below $2$ GeV are also analytically included. This model uses no form factors in the hadronic part of the $\pi N$ interaction. The fit to 
the $\pi N$ measurements in the low-energy region ($T \leq 100$ MeV) involves seven parameters, one of which is the $\pi N$ coupling constant. At this point, it is interesting to examine which of the $\pi N$ coupling 
constants (or their combinations) are determined in the model fits to the $\pi N$ experimental data.

There is no doubt that the fit to the two ES DBs determines $f_c^2$. The nucleon $u$-channel graph in the $\pi^+ p$ case involves the vertices $p \to \pi^+ n$ and $\pi^+ n \to p$. The former is associated with $f_-$, the 
latter with $f_+$; therefore, the $\pi^+ p$ scattering amplitude involves the product $f_- f_+ \equiv f_c^2$. The same applies to $\pi^- p$ ES. On the other hand, the model fits to the CX measurements involve unusual 
combinations of the coupling constants: the $s$-channel graph involves the combination $f_- f_n$, whereas the $u$-channel graph $f_- f_p$.

For over two decades, fits to the ES data have routinely been performed; occasional fits to the CX DB were also attempted, but they were rarely used because of the strong correlations among the model parameters when the 
input DB contains measurements of only one reaction. The origin of these correlations is not difficult to identify. In the fits to the two ES DBs, the $\pi^+ p$ data essentially fix the isospin $I=3/2$ partial-wave 
amplitudes, whereas the $I=1/2$ amplitudes are determined from the $\pi^- p$ ES data. (The $\pi^+ p$ scattering amplitudes are pure $I=3/2$ in nature, whereas the $\pi^- p$ ES amplitudes receive both $I=1/2$ and $I=3/2$ 
contributions.) Of course, the combined fit to the two ES reactions ensures that the finally extracted $I=3/2$ amplitudes have been adjusted (during the optimisation) in such a way as to describe optimally both ES DBs. The 
CX reaction also involves a combination of the two isospin amplitudes (a different one to that of the $\pi^- p$ ES). The exclusive fits to the CX DB cannot reliably determine both isospin amplitudes; measurements of another 
reaction are needed. For over two decades (i.e., between 1990 and 2011), an indirect approach had been followed in the investigation of the violation of the isospin invariance in the low-energy $\pi N$ interaction: the 
scattering amplitudes, obtained from the model fits to the ES DBs, were used in order to \emph{predict} the amplitude of the CX reaction. The reproduction of the CX measurements on the basis of that amplitude was 
subsequently pursued (and was always found very poor). The 1997 report of the isospin-breaking effects in the low-energy $\pi N$ interaction \cite{matsinos1997b} rested upon this approach.

In 2012, a direct approach in the investigation of the violation of the isospin invariance at low energy was implemented. Since then, two types of fits are being performed: the first to the ES measurements, whereas the 
second attempts the simultaneous description of the $\pi^+ p$ and CX DBs. Both isospin amplitudes can be determined in both fits. A comparison between the results of these two fits enables tests of the isospin invariance 
in the $\pi N$ system.

As aforementioned, the combined fit to the $\pi^+ p$ and CX DBs involves both $f_c^2$ (because of the $\pi^+ p$ reaction), as well as the products $f_- f_n$ and $f_- f_p$ (relevant in the case of the CX reaction). Therefore, 
it is not possible to associate the results of this fit with any of the combinations of Section \ref{sec:Definitions}. There is, however, one way in which this analysis is useful. If the isospin invariance is fulfilled in 
the low-energy $\pi N$ interaction, there should be only one $\pi N$ coupling constant, and, regardless of which input DB is used, the fitted values of the model parameter $g_{\pi N N}$ should come out compatible. In fact, 
all analyses since 2012 suggest that $f_{DB0+}>f_c$, where $f_{DB0+}$ is the result of the fit to the combined $\pi^+ p$ and CX DBs. This result is significant and refutes the possibility that all $\pi N$ data for $T \leq 100$ 
MeV could be described with one $\pi N$ coupling constant. Provided that $f_p \approx f_n$ and $f_+ \approx f_-$, all the model analyses of the (combined) $\pi^+ p$ and CX DBs after 2012 suggest that $f_0>f_c$. I will now 
give the $f_c$ and $f_{DB0+}$ values obtained with the ETH model thus far.

The first fits of the ETH model did not involve genuine measurements. During the first years of development and application of the model, fits were made to phase-shift results of various PWAs, e.g., of KH80, of KA85, of 
CMU-LBL, etc. From the phase shifts up to the energy corresponding to the position of the $\Delta(1232)$ resonance, the first fitted value of $g_{\pi N N} = 12.95 \pm 0.08 ({\rm stat.}) \pm 0.03 ({\rm syst.})$ was obtained 
in 1993 \cite{goudsmit1993}. Similar values followed in 1994 \cite{goudsmit1994a,goudsmit1994b}.

The first estimate for $f_c^2$ from genuine measurements was reported in the MENU'97 Conference \cite{matsinos1997a} and later on appeared in the first (in this research programme) study of the violation of the isospin 
invariance in the $\pi N$ system \cite{matsinos1997b}. I will shortly explain why I decided to include only this $f_c^2$ value in Section \ref{sec:DeterminationsOfChargedCoupling}. Additional determinations of $f_c$ were 
made in Refs.~\cite{matsinos2006,matsinos2012,matsinos2013,matsinos2014,matsinos2017a} and of $f_{DB0+}$ in Refs.~\cite{matsinos2013,matsinos2017a}. These recent determinations are all based on the use of the Arndt-Roper 
formula \cite{arndt1972} in the optimisation and, as such, they rest upon the identification (and removal) of the outliers contained in the input DBs. On the other hand, the studies \cite{matsinos1997a,matsinos1997b} 
featured a robust fit to the input data. No data point had to be removed, as the minimisation function had been chosen in such a way as to match the distribution of the normalised residuals: in the framework of 
Refs.~\cite{matsinos1997a,matsinos1997b}, one could argue that there are no outliers; in the more conventional language of the $\chi^2$ minimisation function, one could say that any outliers in the input DB are rendered 
harmless (by applying smaller weights to these points, in comparison to those assigned to the ordinary points). Table \ref{tab:ETH} provides the list of all $f_c^2$ and $f_{DB0+}^2$ values, obtained with the model since 
2006.

\begin{table}
{\bf \caption{\label{tab:ETH}}}Values of the coupling constants $f_c^2$ and $f_{DB0+}^2$ obtained with the ETH model of the $\pi N$ interaction since 2006. The difference between the values of (on one hand) 
Refs.~\cite{matsinos2006,matsinos2012,matsinos2013,matsinos2014} and (on the other) Ref.~\cite{matsinos2017a} is predominantly due to the inclusion of the BERTIN76 DCSs \cite{bertin1976} in the input $\pi^+ p$ DB.
\vspace{0.3cm}
\begin{center}
\begin{tabular}{|c|c|c|}
\hline
Reference & $f_c^2$ & $f_{DB0+}^2$ \\
\hline
\hline
\cite{matsinos2006} & $0.0733(14)$ & $-$ \\
\cite{matsinos2012} & $0.0726(14)$ & $-$ \\
\cite{matsinos2013} & $0.0726(14)$ & $0.0794(13)$ \\
\cite{matsinos2014} & $0.0723(14)$ & $-$ \\
\cite{matsinos2017a} (ZRH17) & $0.0746(14)$ & $0.0805(11)$ \\
\hline
\end{tabular}
\end{center}
\vspace{0.5cm}
\end{table}

`Large' estimates for $f_c^2$ have never been extracted from the model fits to the low-energy $\pi N$ data. The largest $f_c^2$ value, ever obtained, was the one reported in Refs.~\cite{matsinos1997a,matsinos1997b}, when 
the robust fit was carried out and no rescaling of the input data sets was permitted. On the other hand, all $f_{DB0+}^2$ estimates since 2012 have been close to the canonical value.

Regarding my reluctance to use any of the $f_c^2$ values of Table \ref{tab:ETH} in Section \ref{sec:DeterminationsOfChargedCoupling}, one word is due. I consider the $f_c^2$ value, obtained in Refs.~\cite{matsinos1997a,matsinos1997b}, 
to be final in relation to the methodology, as well as to the DB content in the late 1990s. On the other hand, Refs.~\cite{matsinos2006,matsinos2012,matsinos2013,matsinos2014}, though properly published, represent (in my 
opinion) \emph{improvements} in the approach set forward for the identification of the outliers in the input DBs. I believe that the approach was perfected in Ref.~\cite{matsinos2017a}, to the extent that I consider that 
paper as representing the `state-of-the-art' for an optimisation resting upon the use of the Arndt-Roper formula. Had it been properly published, I would have included the $f_c^2$ result of Ref.~\cite{matsinos2017a} in 
Section \ref{sec:DeterminationsOfChargedCoupling}.

\section{\label{sec:Conclusions}Conclusions}

Results for the various $\pi N$ coupling constants were discussed and analysed in this work. Included in the statistical analysis of Section \ref{sec:DeterminationsOfChargedCoupling} were twenty values of the charged-pion 
coupling constant $f_c^2$, namely those of the reported results which fulfil the selection criteria put forward in Section \ref{sec:Introduction}. To the best of my knowledge, only four reports of the neutral-pion coupling 
constant $f_0^2$ may be found in the literature, three of the coupling of the neutral pion to the proton ($f_p^2$), and only one of the coupling of the neutral pion to the neutron ($f_n^2$).

A scatter plot of the values of the charged-pion coupling constant and the year in which they were reported is displayed in Fig.~\ref{fig:ChargedCouplingConstant}. The $f_c^2$ determinations from the $\pi N$ data cluster 
well around their weighted average; on the other hand, those extracted from the $N N$ data exhibit sizeable fluctuation. A representative average, namely $f_c^2 = 762.9^{+6.5}_{-6.2} \cdot 10^{-4}$, was obtained in 
Section \ref{sec:DeterminationsOfChargedCoupling} from robust fits to the $f_c^2$ values of Fig.~\ref{fig:ChargedCouplingConstant}.

Based on the averages of $f_c^2$ and $f_0^2$, given in Sections \ref{sec:DeterminationsOfChargedCoupling} and \ref{sec:DeterminationsOfNeutralCoupling} respectively, there seems to be no evidence that $f_c \neq f_0$. The 
paired test, which is described at the end of Section \ref{sec:DeterminationsOfNeutralCoupling}, yielded inconclusive results: no significant splitting effect was observed in two studies \cite{bugg1995,swart1997}, one study 
reported results marginally compatible with no splitting \cite{navarro2017}, whereas a significant effect was observed in the fourth study \cite{babenko2017}. It is worth mentioning that the analysis of the low-energy $\pi N$ 
data with the ETH model results in significant splitting in the coupling constant $g_{\pi N N}$, see Table \ref{tab:ETH}. However, the effect observed in Ref.~\cite{babenko2017} (namely, $f_c>f_0$) is opposite to the 
results extracted from the $\pi N$ data with the ETH model ($f_c<f_0$) provided that $f_p \approx f_n$ and $f_+ \approx f_-$.

\begin{ack}
I acknowledge a useful discussion with R.G.E.~Timmermans; he drew my attention to the results of Refs.~\cite{stoks1993,timmermans1994,rentmeester1999}, which should be considered to be final in the Nijmegen programme. I 
would like to thank I.I.~Strakovsky for sending me two papers which I had missed, namely Refs.~\cite{arndt2000,pavan2000}, and one which I was aware of (Ref.~\cite{arndt1994b}), yet I had overlooked in the previous version 
of this work. I also acknowledge numerous stimulating discussions over the years with A.~Badertscher and, naturally, with G.~Rasche, in particular regarding the EM corrections in the $\pi N$ system. The use of the hadronic 
masses (instead of the physical ones) in the EM corrections, discussed in Section \ref{sec:FurtherRemarks}, is not one of my own ideas; it is the outcome of a long-term debate between G.~Rasche and W.S.~Woolcock. I am 
obliged to G.~Rasche also for drawing my attention to Refs.~\cite{rasche1976,rasche1982}.

The Feynman graphs of this paper were drawn with the software package JaxoDraw \cite{binosi2004,binosi2009}, available from jaxodraw.sourceforge.net. The remaining figures have been created with MATLAB$^{\textregistered}$~(The 
MathWorks, Inc., Natick, Massachusetts, United States).
\end{ack}

\clearpage
\newpage
\appendix
\section{\label{App:AppA}The EM corrections at the $\pi N$ threshold}

The determination of the $\pi N$ coupling constant from the isovector hadronic scattering length $\tilde{b}_1$, by use of the GMO sum rule, gained momentum during the recent past, in parallel with the remarkable enhancement 
of the low-energy $\pi N$ DB, which the experiments, conducted at the three meson factories (LANL, PSI, and TRIUMF) after 1980, achieved. The analysis of the low-energy DB enables the extraction of reliable estimates for 
the $\pi N$ scattering lengths. In addition, the measurement of the total decay width $\Gamma_{1 s}$ of the ground state of pionic hydrogen in 1995 permitted the direct (i.e., not involving an extrapolation of the $\pi N$ 
scattering amplitudes to threshold) extraction of $b_1$. Of course, EM effects beyond the direct EM contribution need to be removed in both cases, i.e., both from the scattering amplitudes before they can be extrapolated to 
threshold, as well as from $b_1$ obtained from $\Gamma_{1 s}$. I decided to include in this work a rather detailed description of the approaches tailored to the removal of these EM effects from the measurements on pionic 
hydrogen at threshold.

The most recent compilation of the physical constants \cite{pdg2018} has been used in extracting the numerical results below. All masses are expressed in energy units. The uncertainties are total, i.e., they include the 
effects of the variation of all physical `constants' involved, as well as of those relating to the variation of the experimental (and, as far as the EM corrections are concerned, theoretical) input. The results have been 
obtained by means of a Monte-Carlo generation of one billion events. In the calculation, all scattering lengths were expressed exclusively in length units (i.e., fm in this case, not $m_c^{-1}$). However ludicrous I find 
to express lengths in units of inverse mass, I felt somewhat compelled to give some of the resulting scattering lengths also in $m_c^{-1}$ in order to facilitate the comparison with other works.

I will commence with one remark relating to the procedure yielding the strong shift $\epsilon_{1 s}$ of the ground state in pionic hydrogen. What is identified as $\epsilon_{1 s}$ in the two PSI experiments on pionic hydrogen 
is simply the difference between two energy differences. The first of these differences relates to the EM transition energy between the $n p$ and $1 s$ states of pionic hydrogen $E_{n p \to 1 s}^{\rm EM}$, the second to the 
experimentally measured transition energy $E_{n p \to 1 s}$; in fact, this difference is equal to $\epsilon_{1 s}-\epsilon_{n p}$, where $\epsilon_{n p}$ denotes the strong shift of the $n p$ state in pionic hydrogen. To 
the best of my knowledge, the only work which provides estimates for the quantities $\epsilon_{2 p}$ and $\Gamma_{2 p}$ in pionic hydrogen is Ref.~\cite{rasche1982}: therein, it was found that $\epsilon_{2 p}$ is several 
orders of magnitude smaller than $\epsilon_{1 s}$ (see quantities $Re \Delta W (2, p_{1/2})$ and $Re \Delta W (2, p_{3/2})$ on p.~415 of that paper). Therefore, the assumption that $\epsilon_{n p} \approx 0$ makes sense 
$\forall n \geq 2$, and the difference $E_{n p \to 1 s}^{\rm EM} - E_{n p \to 1 s}$ may safely be identified with $\epsilon_{1 s}$.

The first precise $\epsilon_{1 s}$ result, accompanied by a relative accuracy below $1$ \%, was obtained in an experiment conducted at PSI in the mid 1990s \cite{sigg1995,sigg1996b}; the extracted value, obtained using the 
state-of-the-art (at that time) $E_{3 p \to 1 s}^{\rm EM}$, was: $\epsilon_{1 s} = -7.127 \pm 0.028 ({\rm stat.}) \pm 0.036 ({\rm syst.})$ eV. Several improvements in the beamline and in the experimental set-up (for a 
detailed list, see Section 2.2 of Ref.~\cite{schroeder2001}) culminated in $\epsilon_{1 s} = -7.108 \pm 0.013 ({\rm stat.}) \pm 0.034 ({\rm syst.})$ eV, which was the final result of the ETHZ-Neuch{\^a}tel-PSI Collaboration 
\cite{schroeder1999,schroeder2001}.

One decade after Ref.~\cite{schroeder2001} appeared, the $n p \to 1 s$ EM transition energies were re-assessed for $n = 2, 3, 4$ \cite{schlesser2011}. Using the updated $E_{3 p \to 1 s}^{\rm EM}$, the authors of 
Ref.~\cite{schlesser2011} extracted a new $\epsilon_{1 s}$ value from the experimentally measured transition energy $E_{3 p \to 1 s}$ of Ref.~\cite{schroeder2001}; the updated result was: 
$\epsilon_{1 s} = -7.085 \pm 0.013 ({\rm stat.}) \pm 0.034 ({\rm syst.})$ eV. However, the $E_{n p \to 1 s}^{\rm EM}$ estimates were obtained in Ref.~\cite{schlesser2011} on the basis of an earlier $m_c$ value, namely of 
$139.57018(35)$ MeV, as well as of the charge radius of the proton $\sqrt{\avg{r_E^2}}=0.84087(26)(29)$ fm, extracted from muonic hydrogen \cite{antognini2013}. Therefore, one needs to correct $\epsilon_{1 s}$, quoted in Section 
III of Ref.~\cite{schlesser2011}, by using the current $m_c$ value $139.57061(24)$ MeV \cite{pdg2018} and, if one so wishes, by replacing the $\sqrt{\avg{r_E^2}}$ value of Ref.~\cite{antognini2013} by the result $\sqrt{\avg{r_E^2}}=0.8751(61)$ 
fm of the CODATA 2014 compilation \cite{mohr2016} (similar results had been obtained in earlier compilations), obtained from spectroscopic measurements on hydrogen and deuterium, and from $e p$ ES data. Reference 
\cite{pdg2018} simply quotes both $\sqrt{\avg{r_E^2}}$ results, retaining a neutral thesis and making no attempt towards a resolution of the discrepancy between the two values.

At the present time, the EM form factors used in the ETH model \cite{venkat2011} are compatible with the value of the CODATA 2014 compilation \cite{mohr2016}. Self-consistence thus dictates that, in order that the estimates 
for $a_{cc}$ be included in the DB in the analyses involving the ETH model, corrections need to be applied to the $\epsilon_{1 s}$ of Ref.~\cite{schroeder1999,schroeder2001}, as quoted in Section III of Ref.~\cite{schlesser2011}, 
in relation both to the $m_c$ mass, as well as to the use of the $\sqrt{\avg{r_E^2}}$ result of Ref.~\cite{antognini2013} in Ref.~\cite{schlesser2011}. It should be stressed that this choice may not be taken as a preference 
for the $\sqrt{\avg{r_E^2}}$ result of the CODATA 2014 compilation \cite{mohr2016}; it is simply dictated by self-consistence. Similarly, the result of Ref.~\cite{hennebach2014} (which had been obtained on the basis of 
Ref.~\cite{schlesser2011}) needs to be slightly updated.

The effects, induced on $\epsilon_{1 s}$ by the use of the current $m_c$ value and by the use of the $\sqrt{\avg{r_E^2}}$ value of the CODATA 2014 compilation \cite{mohr2016}, are opposite in sign and, to a large extent, 
cancel one another: the former change results in a correction of about $+7.7$ meV, whereas the latter in about $-5.1$ meV. As a result, the net correction is about $2.6$ meV, i.e., about $37$ \% of the statistical uncertainty 
of the $\epsilon_{1 s}$ value in Ref.~\cite{hennebach2014}.

The $\epsilon_{1 s}$ value of the first PSI experiment on pionic hydrogen \cite{sigg1995,sigg1996b,schroeder1999,schroeder2001}, following Ref.~\cite{schlesser2011} along with the updated $m_c$ mass of Ref.~\cite{pdg2018} 
and the use of the $\sqrt{\avg{r_E^2}}$ value of the CODATA 2014 compilation \cite{mohr2016}, should rather read as:
\begin{equation} \label{eq:EQ04_5}
\epsilon_{1 s} = -7.082 \pm 0.013 ({\rm stat.}) \pm 0.034 ({\rm syst.}) \, \, {\rm eV} \, \, \, .
\end{equation}
This value will be used for obtaining the numerical results of Table \ref{tab:EMCorrections}. Regarding the $\epsilon_{1 s}$ value of the Pionic-Hydrogen Collaboration \cite{hennebach2014}, one obtains:
\begin{equation} \label{eq:EQ04_6}
\epsilon_{1 s} = -7.0832 \pm 0.0071 ({\rm stat.}) \pm 0.0064 ({\rm syst.}) \, \, {\rm eV} \, \, \, .
\end{equation}

Introduced by Deser and collaborators \cite{deser1954}, the first of the Deser formulae~\footnote{I prefer this short form as reference to the two important relations developed by Deser, Goldberger, Baumann, and Thirring 
in 1954, as well as by Trueman in 1961, rather than Deser-Goldberger-Baumann-Thirring, Deser-Trueman, or Trueman-Deser relations.} relates $\epsilon_{1 s}$ in pionic hydrogen with the `untreated' (i.e., containing effects 
of EM origin) scattering length $a_{cc}$ which, as already mentioned in Section \ref{sec:ScatteringLengths}, is associated with $\pi^- p$ ES.
\begin{equation} \label{eq:EQ05}
\epsilon_{1 s}=- 4 \frac{E_{1 s}}{r_B} a_{cc} = - \frac{2 \alpha^3 \mu^2}{\hbar c} a_{cc} \, \, \, ,
\end{equation}
where $E_{1 s}=\alpha^2 \mu / 2$ is the (point-Coulomb) EM binding energy of the $1 s$ level and $r_B=\hbar c / (\alpha \mu)$ is the Bohr radius; $\alpha$ denotes the fine-structure constant and $\mu$ stands for the reduced 
mass of the $\pi^- p$ system.

After combining the statistical and systematic uncertainties in Eq.~(\ref{eq:EQ04_5}) linearly~\footnote{The ETHZ-Neuch{\^a}tel-PSI Collaboration favoured the linear combination of their statistical and systematic 
uncertainties, see Section 4.1 of Ref.~\cite{schroeder2001}.}, one obtains from Eq.~(\ref{eq:EQ05}): $a_{cc}=0.12182(81)$ fm.

The second of the Deser formulae, put into its current form by Trueman \cite{trueman1961}, enables the extraction (from $\Gamma_{1 s}$) of the scattering length $a_{c0}$, associated with the $\pi^- p$ CX reaction.
\begin{equation} \label{eq:EQ06}
\Gamma_{1 s}=8 q_0 \frac{E_{1 s}}{r_B \hbar c} \left( 1 + P^{-1} \right) a_{c0}^2 = 4 q_0 \frac{\alpha^3 \mu^2}{(\hbar c)^2} \left( 1 + P^{-1} \right) a_{c0}^2 \, \, \, ,
\end{equation}
where $q_0$ stands for the magnitude of the CM $3$-momentum of the outgoing $\pi^0$ (or neutron) and $P=1.546(9)$ is known as Panofsky ratio. The $\Gamma_{1 s}$ result of Ref.~\cite{schroeder2001} was:
\begin{equation} \label{eq:EQ06_5}
\Gamma_{1 s} = 0.868 \pm 0.040 ({\rm stat.}) \pm 0.038 ({\rm syst.}) \, \, {\rm eV} \, \, \, ,
\end{equation}
From Eqs.~(\ref{eq:EQ06},\ref{eq:EQ06_5}), one obtains~\footnote{The scattering length $a_{c0}$ is negative.} $a_{c0}=-0.1784(81)$ fm.

Evidently, the ETHZ-Neuch{\^a}tel-PSI Collaboration delivered $a_{cc}$ to an accuracy of $0.66$ \% and $a_{c0}$ to an accuracy of $4.5$ \%. At this point, corrections need to be applied, in order to rid $a_{cc}$ and $a_{c0}$ 
of effects of EM origin (i.e., of the interference effects between the hadronic potential and the potentials associated with the vacuum polarisation, the extended-charge distributions of the pion and of the proton, etc.), 
and lead to estimates for the corresponding hadronic scattering lengths, $\tilde{a}_{cc}$ and $\tilde{a}_{c0}$. The EM corrections are usually expressed in the form of two quantities, $\delta_\epsilon$ for $a_{cc}$ and 
$\delta_\Gamma$ for $a_{c0}$. The two hadronic scattering lengths are obtained from $a_{cc}$ and $a_{c0}$ according to the following two definitions.
\begin{equation} \label{eq:EQ07}
\tilde{a}_{cc} = a_{cc} / (1 + \delta_\epsilon)
\end{equation}
\begin{equation} \label{eq:EQ08}
\tilde{a}_{c0} = a_{c0} / (1 + \delta_\Gamma)
\end{equation}

It is understood that the scattering lengths $a_{cc}$ and $a_{c0}$ in Eqs.~(\ref{eq:EQ07},\ref{eq:EQ08}) are associated with the original Deser formulae (\ref{eq:EQ05},\ref{eq:EQ06}). These formulae represent \emph{leading-order} 
(LO) evaluations of $\epsilon_{1 s}$ and $\Gamma_{1 s}$, namely evaluations at $\mathcal{O}(\alpha^3)$. In several works, the quantities $a_{cc}$ and $a_{c0}$ of Eqs.~(\ref{eq:EQ05},\ref{eq:EQ06}) are therefore denoted as 
$a_{cc}^{\rm LO}$ and $a_{c0}^{\rm LO}$. This is done in order to distinguish these scattering lengths from those appearing in the upgraded forms of the Deser formulae, i.e., the expressions obtained at higher orders of 
$\alpha$. At the present time, only the next-to-leading-order (NLO) evaluations of $\epsilon_{1 s}$ and $\Gamma_{1 s}$ are available, i.e., the evaluations at $\mathcal{O}(\alpha^4)$. Some authors denote the scattering 
lengths, entering the NLO evaluations of $\epsilon_{1 s}$ and $\Gamma_{1 s}$, as $a_{cc}^{\rm NLO}$ and $a_{c0}^{\rm NLO}$. In this work, $a_{cc}$ and $a_{c0}$ will represent $a_{cc}^{\rm LO}$ and $a_{c0}^{\rm LO}$, 
respectively. I will introduce $\mathcal{A}_{cc}$ and $\mathcal{A}_{c0}$ later on, to make reference to $a_{cc}^{\rm NLO}$ and $a_{c0}^{\rm NLO}$, respectively.

Regarding the removal of the EM effects at threshold before the first experiment on pionic hydrogen was conducted at PSI in the mid 1990s, I am only aware of two papers by Rasche and Woolcock \cite{rasche1976,rasche1982}. 
The former paper develops the methodology needed for the correct inclusion of the effects of the $\gamma n$ channel at threshold. The second paper presents a method for the determination of the strong shift and total decay 
width in pionic atoms. The numerical results in Section 3 of Ref.~\cite{rasche1982} were tailored to the $2 p \to 1 s$ transition in pionic hydrogen, which was deemed at the time as the most promising transition (on the 
basis of the yield). Comments on the corrections of Ref.~\cite{rasche1982} may be found in Section 4 of Ref.~\cite{sigg1996a}. The corrections of Ref.~\cite{rasche1982} were superseded by those of Ref.~\cite{oades2007}.

Several schemes of EM corrections were developed after the first experimental results at threshold became available. Some of these correction schemes aim at the removal of the `trivial' EM effects, i.e., of those associated 
with the contributions from the vacuum polarisation, from the extended-charge distributions of the pion and of the proton, as well as from the mass differences of the particles in the initial and final states. In the context 
of Ref.~\cite{oades2007}, all these contributions comprise the stage-1 EM corrections. The models of Sections \ref{sec:PotentialModels} and \ref{sec:ELWModel} are expected to remove these effects. On the other hand, works 
carried out within the framework of Chiral Perturbation Theory (ChPT) also attempt the removal of the effects which are associated with the mass difference between the $u$ and $d$ quarks, i.e., effects which belong to the 
stage-2 EM corrections in the context of Ref.~\cite{oades2007}. The models of Sections \ref{sec:Lyubovitskij2000}, \ref{sec:Gasser2002}, \ref{sec:Zemp2004}, and \ref{sec:Meissner} belong to this category. When comparing 
the results of the various correction schemes, one needs to bear in mind the distinction between these two categories of corrections.

\subsection{\label{sec:PotentialModels}Potential models for the removal of the EM effects}

In their assessment of the EM effects at threshold, Refs.~\cite{sigg1996a,oades2007} made use of suitable potentials.

An estimate for $\delta_\epsilon$ was obtained in Ref.~\cite{sigg1996a} by means of a two-channel calculation, along with the phenomenological addition of the effects of the $\gamma n$ channel: $\delta_\epsilon=-2.1(5) \cdot 10^{-2}$. 
This $\delta_\epsilon$ value yields $\tilde{a}_{cc}=0.1244(10)$ fm or, for those who prefer to express the scattering lengths in $m_c^{-1}$, $\tilde{a}_{cc}=0.08801(74) \, m_c^{-1}$.

In Ref.~\cite{oades2007}, the correction $\Delta a_{cc} \coloneqq a_{cc} - \tilde{a}_{cc}$ was evaluated by means of a three-channel calculation: $\Delta a_{cc}=0.0008(8)$ fm. Applying this correction to the untreated 
$a_{cc}$ value, emerging from the updated $\epsilon_{1 s}$ result of Ref.~\cite{schroeder2001}, leads to $\tilde{a}_{cc}=0.1210(11)$ fm or $0.08560(80) \, m_c^{-1}$. To enable the comparison of the corrections, obtained 
in the two papers, one may translate the $\Delta a_{cc}$ value of Ref.~\cite{oades2007} into a $\delta_\epsilon$ value; one obtains $\delta_\epsilon=0.67(67) \cdot 10^{-2}$.

In Ref.~\cite{sigg1996a}, the estimate for $\delta_\Gamma$ of $-1.3(5) \cdot 10^{-2}$ had been extracted. On the other hand, the correction of Ref.~\cite{oades2007} had been expressed as the difference between the untreated 
and the corrected scattering lengths for the CX reaction: $\Delta a_{c0} \coloneqq a_{c0} - \tilde{a}_{c0}$. Expressed as a $\delta_\Gamma$ value, the correction $\Delta a_{c0}$ of Ref.~\cite{oades2007} would have been: 
$\delta_\Gamma=-1.66(33) \cdot 10^{-2}$. One therefore concludes that the two corrections \cite{sigg1996a,oades2007} agree within the uncertainties in the case of $\Gamma_{1 s}$. Evidently, only the correction applied to 
$a_{cc}$ is sensitive to the treatment of the $\gamma n$ channel. The $\Gamma_{1 s}$ result of Ref.~\cite{schroeder2001}, along with the EM correction $\Delta a_{c0}$ of Ref.~\cite{oades2007}, yields $\tilde{a}_{c0}=-0.1814(81)$ 
fm or $-0.1283(57) \, m_c^{-1}$. The $\tilde{a}_{c0}$ result of Ref.~\cite{schroeder2001}) (i.e., $-0.128(6) \, m_c^{-1}$, see Eq.~(30) therein) was identical in the physical sense.

\subsection{\label{sec:ELWModel}The model of Ericson, Loiseau, and Wycech \cite{ericson2004}}

In 2004, Ericson and collaborators \cite{ericson2004} followed a non-relativistic approach using Coulomb wavefunctions, with a short-range hadronic interaction and extended-charge distributions, and treated four sources of 
EM corrections: the first two originate from the interference of two potentials (vacuum polarisation, extended-charge distributions) with the hadronic potential; the remaining corrections relate to renormalisation and gauge 
contributions.

The estimates of Ref.~\cite{ericson2004} for the corrections $\delta_\epsilon$ and $\delta_\Gamma$ may be found in their Table 1: $\delta_\epsilon = -0.62(29) \cdot 10^{-2}$ and $\delta_\Gamma = 1.02(23) \cdot 10^{-2}$. 
The discrepancy in the $\delta_\Gamma$ values between the results of Sections \ref{sec:PotentialModels} and \ref{sec:ELWModel} is noticeable. The correction $\delta_\epsilon$ of Ref.~\cite{ericson2004} lies in-between the 
results obtained with the two potential models of the previous section, slightly closer to the result of Ref.~\cite{oades2007}.

Comments on the approach of Ref.~\cite{ericson2004} may be found in Section 4 of Ref.~\cite{oades2007}; there is no point in repeating them here. I will only mention that the uncertainties of $\delta_\epsilon$ and 
$\delta_\Gamma$ of Ref.~\cite{ericson2004} appear to be optimistic.

\subsection{\label{sec:Lyubovitskij2000}The Lyubovitskij-Rusetsky correction to $a_{cc}$ \cite{lyub2000}}

I consider the 2000 paper of Lyubovitskij and Rusetsky \cite{lyub2000} important for two reasons.
\begin{itemize}
\item The authors presented an evaluation of $\epsilon_{1 s}$ at $\mathcal{O}(\alpha^4)$; this is an essential upgrade of Eq.~(\ref{eq:EQ05}). Part of the effects, which need to be taken care of by the EM corrections in 
case that Eq.~(\ref{eq:EQ05}) is used, are contained in the upgraded expression.
\item Their work constituted the first attempt to derive the EM corrections within the (systematic) framework of ChPT.
\end{itemize}

The relation at $\mathcal{O}(\alpha^4)$ between $\epsilon_{1 s}$ and the scattering length (denoted as $\mathcal{A}$ in Ref.~\cite{lyub2000}, $\mathcal{A}_{cc}$ in this work) reads as:
\begin{equation} \label{eq:EQ09}
\epsilon_{1 s} = - \frac{2 \alpha^3 \mu^2}{\hbar c} \mathcal{A}_{cc} \left( 1 + 2 \alpha \left( 1 - \ln \alpha \right) \frac{\mu \mathcal{A}_{cc}}{\hbar c} \right) \, \, \, ,
\end{equation}
which, after appending the effects due to the interference between the strong interaction and the vacuum polarisation $\varphi \approx 0.483 \cdot 10^{-2}$ of Ref.~\cite{eiras2000} (these effects had not been included in 
Ref.~\cite{lyub2000}), may be rewritten as
\begin{equation} \label{eq:EQ10}
\epsilon_{1 s} = - \frac{2 \alpha^3 \mu^2}{\hbar c} \mathcal{A}_{cc} \left( 1 + \varphi + 2 \alpha \left( 1 - \ln \alpha \right) \frac{\mu \mathcal{A}_{cc}}{\hbar c} \right) \, \, \, .
\end{equation}
Lyubovitskij and Rusetsky did not identify $\mathcal{A}_{cc}$ with $\tilde{a}_{cc}$. Additional isospin-breaking corrections (denoted as $\epsilon$ in Ref.~\cite{lyub2000}, $\Delta \mathcal{A}_{cc}$ in this work), to be 
understood as residual effects of EM origin and contributions originating from the mass difference between the $u$ and $d$ quarks, were evaluated in Ref.~\cite{lyub2000} at $\mathcal{O}(p^2)$ in ChPT. The relation between 
the quantities $\mathcal{A}_{cc}$ and $\tilde{a}_{cc}$ was given in Ref.~\cite{lyub2000} as:
\begin{equation*}
\tilde{a}_{cc} = \mathcal{A}_{cc} - \Delta \mathcal{A}_{cc} \, \, \, .
\end{equation*}
The correction $\Delta \mathcal{A}_{cc}$ depends on three low-energy constants (LECs) $c_1$, $f_1$, and $f_2$ in the $\mathcal{O}(p^2)$ chiral $\pi N$ Lagrangian, one of which ($f_1$) is poorly known. According to 
Ref.~\cite{lyub2000}:
\begin{equation} \label{eq:EQ11}
\Delta \mathcal{A}_{cc} = \frac{m_p \hbar c}{2 (m_p+m_c)} \left( \frac{2 (m_c^2-m_0^2)}{\pi F_\pi^2} c_1 - \alpha (4 f_1 + f_2) \right) \, \, \, ,
\end{equation}
where $m_0$ is the mass of the neutral pion and $F_\pi=92.28(12)$ MeV is the pion-decay constant.
\begin{itemize}
\item For $c_1$, Lyubovitskij and Rusetsky used a result from one of Karlsruhe analyses of the mid 1980s, privately communicated to the authors; that value was equal to $-0.925$ GeV$^{-1}$ (no uncertainty was quoted in 
Ref.~\cite{lyub2000}). Also using information from the Karlsruhe programme of the 1980s \cite{hoehler1983}, Gasser and collaborators \cite{gasser2002} came up (in 2002) with $c_1 = -0.93(7)$ GeV$^{-1}$. In the same year, 
Lyubovitskij and collaborators \cite{lyub2002} imported (from a work of 2001) a different $c_1$ value, namely $c_1=-1.2(1)$ GeV$^{-1}$. More recent works \cite{baru2011a,baru2011b} recommend: $c_1 = -1.0(3)$ GeV$^{-1}$.
\item Regarding the LEC $f_2$, Ref.~\cite{lyub2000} used $f_2 = - 0.97(38)$ GeV$^{-1}$, which is also the recommended value in Ref.~\cite{gasser2002}.
\item As aforementioned, the LEC $f_1$ is poorly known. To derive an estimate for the correction $\delta_\epsilon$, Lyubovitskij and Rusetsky assumed in Ref.~\cite{lyub2000} that $\lvert f_1 \rvert \leq \lvert f_2 \rvert$. 
However, Lyubovitskij and collaborators \cite{lyub2002} arrived in 2002 at a mismatching result for the ratio $f_1/f_2$, namely $2.24(26)$. The authors also favoured $f_1=-2.29(19)$ GeV$^{-1}$, which does not seem to be in 
line with the other `expectations' for this LEC. Gasser and collaborators \cite{gasser2002} mention their ``order of magnitude'' estimate for $\lvert f_1 \rvert$ at about $1.4$ GeV$^{-1}$.
\end{itemize}
Be that as it may, Ref.~\cite{lyub2000} reported a large negative correction: $\delta_\epsilon=(-4.8 \pm 2.0) \cdot 10^{-2}$, where the uncertainty is dominated by the poor knowledge of $f_1$. This correction rests upon the 
assumption $\lvert f_1 \rvert \leq \lvert f_2 \rvert$. I set out to re-evaluate the correction $\delta_\epsilon$ in the Lyubovitskij-Rusetsky scheme, using Eq.~(\ref{eq:EQ10}), rather than Eq.~(\ref{eq:EQ09}) which the 
authors had used. As Ref.~\cite{lyub2000} mentions no uncertainty in the LEC $c_1$, I first assumed that $c_1$ was not varied in their analysis; however, the resulting uncertainty of $\delta_\epsilon$ turned out to be nearly 
a factor of $2$ smaller than the one quoted in Ref.~\cite{lyub2000}. Therefore, I concluded that also $c_1$ was varied in Ref.~\cite{lyub2000} and proceeded by changing the assigned $c_1$ uncertainty, until the final result 
for $\delta_\epsilon$ matched the reported $\delta_\epsilon$ uncertainty of Ref.~\cite{lyub2000}. My conclusion is that Lyubovitskij and Rusetsky had most likely used a $\delta c_1$ value between $0.2$ and $0.3$ GeV$^{-1}$ 
in their work. In any case, the $\delta c_1$ value of $0.3$ GeV$^{-1}$, also recommended in Refs.~\cite{baru2011a,baru2011b}, appears to be reasonable and conservative. For the needs of Table \ref{tab:EMCorrections}, I 
obtain the correction $\delta_\epsilon$ using Eq.~(\ref{eq:EQ10}) with the $\varphi$ value of Ref.~\cite{eiras2000} and $\delta c_1 = 0.3$ GeV$^{-1}$. The other two LECs are varied according to Ref.~\cite{lyub2000}.

As the models of Sections \ref{sec:PotentialModels} and \ref{sec:ELWModel} do not contain any stage-2 EM corrections, the comparison of their $\delta_\epsilon$ values with the result of this section does not make much sense. 
On the other hand, one could pose the question whether a comparison could be meaningful if $\tilde{a}_{cc}$ were identified as the scattering length $\mathcal{A}_{cc}$, obtained from Eq.~(\ref{eq:EQ10}). There is no doubt 
that some of the effects, which are treated by the models of Sections \ref{sec:PotentialModels} and \ref{sec:ELWModel}, are contained in $\Delta \mathcal{A}_{cc}$ of Eq.~(\ref{eq:EQ11}). Unfortunately, because of the mixed 
term in Eq.~(\ref{eq:EQ11}) (last term within the large brackets), one cannot disentangle the EM contributions and those relating to the $m_u \neq m_d$ effects. It appears to me that there is no guarantee that a comparison 
of the results of this section with those obtained with the models of Sections \ref{sec:PotentialModels} and \ref{sec:ELWModel} is meaningful. Nevertheless, I will also obtain a $\delta_\epsilon$ value corresponding to the 
case that $\mathcal{A}_{cc}$ of Eq.~(\ref{eq:EQ10}) is identified as $\tilde{a}_{cc}$. This \emph{intermediate} result will be helpful later on in assessing the importance of the isospin-breaking effects 
$\Delta \mathcal{A}_{cc}$ of Eq.~(\ref{eq:EQ11}).

In 2002, Lyubovitskij and collaborators \cite{lyub2002} provided an update of $\delta_\epsilon$, on the basis of improved knowledge of $f_1$ when employing their ``perturbative chiral quark model''; the new value was 
$\delta_\epsilon=-2.8 \cdot 10^{-2}$, quoted in Ref.~\cite{lyub2002} without an uncertainty. Evidently, the updated value of Ref.~\cite{lyub2002} is not incompatible with the 1996 result extracted with the potential model 
of Ref.~\cite{sigg1996a}.

\subsection{\label{sec:Gasser2002}Isospin-breaking corrections evaluated at $\mathcal{O}(p^3)$ in ChPT \cite{gasser2002}}

An even larger (and more uncertain) correction $\delta_\epsilon$ was extracted in 2002 \cite{gasser2002} within a calculation at NLO ($\mathcal{O}(p^3)$) in isospin breaking and in the low-energy expansion: 
$(-7.2 \pm 2.9) \cdot 10^{-2}$.

\subsection{\label{sec:Zemp2004}Leading-order correction $\delta_\Gamma$ in ChPT \cite{zemp2004}}

The LO correction to $a_{c0}$, derived in ChPT in Ref.~\cite{zemp2004} in 2004, was found small: $\delta_\Gamma=0.6(2) \cdot 10^{-2}$, see Eq.~(5.26) therein. One notices that the correction $\delta_\Gamma$ from ChPT is 
more accurate than the correction $\delta_\epsilon$. This is due to the fact that the LEC $f_1$ does not enter the determination of $\delta_\Gamma$.

\subsection{\label{sec:Meissner}The corrections developed by the Bonn-J{\"u}lich group}

Between 2005 and 2011, the Bonn-J{\"u}lich group developed a correction scheme for the pionic-hydrogen measurements, similar to those detailed in Sections \ref{sec:Lyubovitskij2000}, \ref{sec:Gasser2002}, and \ref{sec:Zemp2004}, 
see Refs.~\cite{baru2011a,baru2011b} and the relevant papers therein. In addition, corrections for the strong shift of $1 s$ state in pionic deuterium were developed.

Regarding $\epsilon_{1 s}$ in pionic hydrogen, Ref.~\cite{baru2011a} uses Eq.~(\ref{eq:EQ10}) to extract $\mathcal{A}_{cc}$, which the authors call $a_{\pi^- p}$ in their paper. They subsequently associate $\mathcal{A}_{cc}$ 
with the difference $b_0 - \tilde{b}_1$.
\begin{equation} \label{eq:EQ12}
b_0 - \tilde{b}_1 = \mathcal{A}_{cc} - \Delta a_{cc} \hbar c \, \, \, ,
\end{equation}
where the isoscalar scattering length is to be thought of as untreated~\footnote{The untreated isoscalar scattering length $b_0$ also enters the strong shift $\epsilon_{1 s}$ in pionic deuterium. A combined analysis of the 
$\epsilon_{1 s}$ values in pionic hydrogen and deuterium, and of $\Gamma_{1 s}$ of pionic hydrogen enables a more accurate determination of the quantities $b_0$ and $\tilde{b}_1$ (compared to the use of the information 
extracted only from pionic hydrogen), see Fig.~2 of Ref.~\cite{baru2011a}.}, as the lack of the tilde over it indicates, and $\Delta a_{cc} = (-2.0 \pm 1.3) \cdot 10^{-3} \, m_c^{-1}$.

For the relation between $\Gamma_{1 s}$ of pionic hydrogen and the corresponding scattering length $\mathcal{A}_{c0}$, the authors use the expression:
\begin{align} \label{eq:EQ13}
\Gamma_{1 s}=4 q_0 \frac{\alpha^3 \mu^2}{(\hbar c)^2} \left( 1 + P^{-1} \right) \mathcal{A}_{c0}^2 \Big( & 1 + \varphi + 4 \alpha \left( 1 - \ln \alpha \right) \frac{\mu \mathcal{A}_{cc}}{\hbar c}\nonumber\\
& + 2 ( m_p + m_c - m_n - m_0 ) \frac{\mu b_0^2}{(\hbar c)^2} \Big) \, \, \, ,
\end{align}
where $m_n$ is the mass of the neutron and $\mathcal{A}_{c0} = \sqrt{2} \tilde{b}_1 + \Delta a_{c0} \hbar c$, with $\Delta a_{c0}=0.4(9) \cdot 10^{-3} \, m_c^{-1}$. Equations (\ref{eq:EQ12},\ref{eq:EQ13}) contain two 
unknowns: $b_0$ and $\tilde{b}_1$. The quantity $\tilde{b}_1$ may be obtained by use of a simple recursion scheme; the convergence is very fast. The quantity $b_0$ is subsequently obtained via Eq.~(\ref{eq:EQ12}). Evident 
from Refs.~\cite{baru2011a,baru2011b} is that the isospin-breaking effects have a larger impact on the isoscalar part of the $\pi N$ interaction at threshold; for this correction, the authors give the expression:
\begin{equation*}
\tilde{b}_0 = b_0 - \frac{m_p \hbar c}{m_p+m_c} \left( \frac{m_c^2 - m_0^2}{\pi F_\pi^2} c_1 - 2 \alpha f_1 \right) \, \, \, ,
\end{equation*}
where the values and uncertainties of the LECs $c_1$ and $f_1$, used in Refs.~\cite{baru2011a,baru2011b}, have already been given in Section \ref{sec:Lyubovitskij2000}. Comparison with Eq.~(\ref{eq:EQ11}) implies that the 
correction to $b_1$ reads as
\begin{equation*}
b_1 - \tilde{b}_1 = \frac{m_p \hbar c}{m_p+m_c} \frac{\alpha f_2}{2} \, \, \, ,
\end{equation*}
and comes out equal to $-0.43(17) \cdot 10^{-3} \, m_c^{-1}$. Presumably, this correction is contained in $\Delta a_{cc}$ of Eq.~(\ref{eq:EQ12}). The corrected $\tilde{a}_{cc}$ may then be obtained as the difference 
$\tilde{b}_0-\tilde{b}_1$, whereas $\tilde{a}_{c0}=\sqrt{2} \tilde{b}_1$.

\subsection{\label{sec:FurtherRemarks}A few remarks on the EM corrections at threshold}

The important results of the application of the aforementioned correction schemes to $\epsilon_{1 s}$ and $\Gamma_{1 s}$ in the case of pionic hydrogen \cite{schroeder2001} are given in Table \ref{tab:EMCorrections}. The 
$f_c^2$ values, contained in the last column of the table, have been obtained by use of the GMO sum rule. The $f_c^2$ value of this work after applying the corrections of Ref.~\cite{ericson2004} is smaller than the value 
quoted in Ref.~\cite{ericson2004} and is accompanied by a considerably larger uncertainty. These changes are due to three reasons: a) The constraint from $\epsilon_{1 s}$ in pionic deuterium had also been used in 
Ref.~\cite{ericson2004}, in order to restrict their estimate for $\tilde{b}_1$; b) Reference \cite{ericson2004} combined the statistical and systematic uncertainties of Ref.~\cite{schroeder2001} quadratically; c) The values 
of the integral $J^-$, used in the GMO sum rule, differ between the two works: Ref.~\cite{ericson2004} uses the estimate of Ref.~\cite{ericson2002}, whereas this work uses the weighted average of three results, one of which 
is the estimate of Ref.~\cite{ericson2002}, see Appendix \ref{App:AppB}. To enable a comparison with the results obtained with the models of Sections \ref{sec:PotentialModels} and \ref{sec:ELWModel}, and also provide an 
impression of the largeness of the $\mathcal{O}(p^2)$ corrections in Ref.~\cite{lyub2000}, a $\delta_\epsilon$ result was obtained after identifying $\mathcal{A}_{cc}$ with $\tilde{a}_{cc}$ or, equivalently, after ignoring 
the correction $\Delta \mathcal{A}_{cc}$ of Eq.~(\ref{eq:EQ11}). The difference between the corrections $\delta_\epsilon$ between Ref.~\cite{lyub2000} and the value of Table \ref{tab:EMCorrections} is accounted for by the 
use of Eq.~(\ref{eq:EQ10}), instead of Eq.~(\ref{eq:EQ09}) which had been used in Ref.~\cite{lyub2000}.

\begin{table}
{\bf \caption{\label{tab:EMCorrections}}}The important results of the application of a few correction schemes to the measurements of $\epsilon_{1 s}$ of Eq.~(\ref{eq:EQ04_5}) and $\Gamma_{1 s}$ of Eq.~(\ref{eq:EQ06_5}) of 
pionic hydrogen \cite{schroeder2001}. The input, common in all cases, comprises the $a_{cc}$ and $a_{c0}$ results obtained using Eqs.~(\ref{eq:EQ05},\ref{eq:EQ06}). All corrections have been expressed in the form 
$\delta_\epsilon$ and $\delta_\Gamma$, see Eqs.~(\ref{eq:EQ07},\ref{eq:EQ08}). The $f_c^2$ values of the last column have been obtained by use of the GMO sum rule, see Appendix \ref{App:AppB}. A $\delta_\epsilon$ result 
was also obtained after identifying the solution $\mathcal{A}_{cc}$ of Eq.~(\ref{eq:EQ10}) with $\tilde{a}_{cc}$.
\vspace{0.3cm}
\begin{center}
\begin{tabular}{|c|c|c|c|c|c|}
\hline
Source & $\delta_\epsilon$ ($10^{-2}$) & $\delta_\Gamma$ ($10^{-2}$) & $\tilde{a}_{cc}$ (fm) & $\tilde{a}_{c0}$ (fm) & $f_c^2$ \\
\hline
\hline
\cite{sigg1996a} & $-2.1(5)$ & $-1.3(5)$ & $0.1244(10)$ & $-0.1808(82)$ & $0.0780(25)$ \\
\cite{ericson2004} & $-0.62(29)$ & $1.02(23)$ & $0.12258(89)$ & $-0.1766(80)$ & $0.0768(24)$ \\
\cite{oades2007} & $0.67(67)$ & $-1.66(33)$ & $0.1210(11)$ & $-0.1814(81)$ & $0.0782(24)$ \\
\hline
\cite{lyub2000}, $\tilde{a}_{cc} \equiv \mathcal{A}_{cc}$ & $1.1234(42)$ & $-$ & $-$ & $-$ & $-$ \\
\hline
\cite{lyub2000} & $-4.3 \pm 2.2$ & $-$ & $0.1273(30)$ & $-$ & $-$ \\
\cite{gasser2002} & $-7.2 \pm 2.9$ & $-$ & $0.1314(42)$ & $-$ & $-$ \\
\cite{zemp2004} & $-$ & $0.6(2)$ & $-$ & $-0.1774(80)$ & $0.0770(24)$ \\
\cite{baru2011a,baru2011b} & $-7.2 \pm 2.6$ & $0.56(72)$ & $0.1314(37)$ & $-0.1774(81)$ & $0.0770(24)$ \\
\hline
\end{tabular}
\end{center}
\vspace{0.5cm}
\end{table}

Visual inspection of Table \ref{tab:EMCorrections} leads to the following conclusions.
\begin{itemize}
\item A consistent picture for the corrections $\delta_\epsilon$ and $\delta_\Gamma$ does not emerge from this table.
\item One may argue that the three-channel calculation of Ref.~\cite{oades2007} constitutes an improvement over the two-channel evaluation of Ref.~\cite{sigg1996a}, and thus proceed to compare the $\delta_\epsilon$ and 
$\delta_\Gamma$ results of Ref.~\cite{oades2007} with those obtained with the only other approach which does not deploy ChPT, namely Ref.~\cite{ericson2004}. Obviously, there is no matching; the signs are opposite in both 
corrections $\delta_\epsilon$ and $\delta_\Gamma$. Moreover, the difference between the two corrections $\delta_\Gamma$ is disturbing.
\item The correction $\delta_\epsilon$ of Ref.~\cite{oades2007} appears compatible with the result obtained from the upgraded Deser formula for $\epsilon_{1 s}$ (see Eq.~(\ref{eq:EQ10})), whereas the corresponding result of 
Ref.~\cite{ericson2004} is not. However, it is not clear that such a comparison is meaningful. Part of the EM corrections of Refs.~\cite{ericson2004,oades2007} are contained in the upgraded Deser formula for $\epsilon_{1 s}$; 
another part is contained in the correction $\Delta \mathcal{A}_{cc}$; a third part is not contained in the correction $\Delta \mathcal{A}_{cc}$. In addition, the correction $\Delta \mathcal{A}_{cc}$ contains effects which 
go beyond those tackled in Refs.~\cite{sigg1996a,ericson2004,oades2007}, e.g., effects emanating from the mass difference between the $u$ and $d$ quarks. Therefore, the compatibility between the correction $\delta_\epsilon$ 
of Ref.~\cite{oades2007} with the result obtained from the upgraded Deser formula for $\epsilon_{1 s}$ could be coincidental.
\item The correction $\delta_\Gamma$ extracted in Ref.~\cite{ericson2004} is compatible with the two estimates obtained within the framework of ChPT in Refs.~\cite{zemp2004,baru2011a,baru2011b}. It has been suggested that 
potential models are prone to yield negative corrections $\delta_\Gamma$. Considering the outcome of Refs.~\cite{sigg1996a,oades2007}, this might indeed be the case.
\item It is time I discussed the corrections obtained within the framework of ChPT. Compared to the experimental uncertainty of $\epsilon_{1 s}$, the corrections $\delta_\epsilon$ of Refs.~\cite{lyub2000,gasser2002,baru2011a,baru2011b} 
are large and, even worse, poorly known. The large uncertainties are attributable to the poor knowledge of the LEC $f_1$. The essential difference between Refs.~\cite{lyub2000,gasser2002} is that, in the former work, the 
additional isospin-breaking effects are treated at $\mathcal{O}(p^2)$; in Ref.~\cite{gasser2002}, they are treated at $\mathcal{O}(p^3)$. If, as the result of the application of the correction $\Delta \mathcal{A}_{cc}$ of 
Eq.~(\ref{eq:EQ11}), $\delta_\epsilon$ changes by as much as $-5.4$ \% (i.e., from $+1.1$ \% to $-4.3$ \%) and the result of the correction at the next order brings another $-2.9$ \%, then I do wonder what surprises the 
calculation at $\mathcal{O}(p^4)$ could bring. If any convergence can be substantiated on the basis of these numbers, then it ought to be a weak one. Moreover, I hardly see a meaningful use of a procedure which increases 
the uncertainty of the correction $\delta_\epsilon$ at `every next order' by $1$ \%. As a result, I cannot understand why the corrections of Refs.~\cite{gasser2002,baru2011a,baru2011b} are applied unquestionably to 
(approximately) twenty times more accurate experimental results as, for instance, the case has been in Ref.~\cite{hennebach2014}.
\item The only positive conclusion from the visual inspection of Table \ref{tab:EMCorrections} is the overall agreement of Refs.~\cite{ericson2004,zemp2004,baru2011a,baru2011b} regarding the correction $\delta_\Gamma$; 
they all agree that this correction is small, below the $1$ \% level. I honestly do not see much else worthy of remembrance in Table \ref{tab:EMCorrections}.
\end{itemize}
In order that the $\delta_\epsilon$ value, obtained from Refs.~\cite{baru2011a,baru2011b}, approach the results extracted with the models of Sections \ref{sec:PotentialModels} and \ref{sec:ELWModel}, the LEC $f_1$ needs 
to be substantially more negative than it is currently allowed ($-1.4$ GeV$^{-1}$). An $f_1$ value in the vicinity of $-4$ GeV$^{-1}$ would lead to vanishing $\delta_\epsilon$.

At this point, I feel that I need to make one statement. Between 1990 and 1995, I had heard at least four prominent theorists lamenting the lack of precise experimental information at threshold. The ETHZ-Neuch{\^a}tel-PSI 
Collaboration provided $\epsilon_{1 s}$ to an accuracy well below $1$ \%, whereas both statistical and systematic uncertainties, reported by the Pionic-Hydrogen Collaboration, are at the level of or below $0.1$ \%. Such 
accuracy is unprecedented in Pion Physics. After this precise information became available, the theorists discovered that no competitive correction scheme had been developed~\footnote{In fact, the lack of such a correction 
scheme served as motivation for Sigg and collaborators to set out to investigate the EM effects in pionic hydrogen in Ref.~\cite{sigg1996a}.} to enable the extraction of the useful hadronic information from the experimental 
results. If the best Theory can do is to provide EM corrections at threshold which are (at least) one order of magnitude less accurate than the experimental results, then my opinion is that Theory needs to find a way to 
catch up.

I left one subtle subject for the end of this section. The way I understand the issue of the EM corrections is as follows. If complete, an EM correction to a value of a physical quantity in this Universe would translate it 
into the corresponding value in a Universe where there is no EM interaction (that Universe will be named `hypothetical'). All available correction schemes aim at the removal of effects relating to the interaction of the 
particles involved, but assume that no change is induced on the particles themselves as the result of the absence of the EM interaction. One may pose the question: What happens to the particle `proton' itself when the EM 
interaction is switched off? The physical mass of the proton surely receives EM contributions; these contributions need to be subtracted in the hypothetical Universe. Therefore, the particle `proton' of this Universe will 
have another mass in the hypothetical Universe (and, of course, will be neutral). The same applies to all other (charged or composite) particles, e.g., to the `neutron' and to the `pions'. If the strong interaction does 
not distinguish between the members of one isospin multiplet, then the hadronic mass of protons and neutrons should be the same. The same applies to the charged and neutral pions, which should share one hadronic mass. 
Therefore, the four physical masses of this Universe (proton, neutron, charged pion, and neutral pion) would reduce to two hadronic masses in the hypothetical Universe, namely the hadronic mass of the nucleon and that of 
the pion. The former should be smaller than the physical mass of the proton, whereas the latter should be smaller than or equal to the physical mass of the neutral pion. Some would suggest that the hadronic mass of the pion 
should be taken to be the physical mass of the neutral pion. That would be a breakthrough (as one hadronic mass in the $\pi N$ interaction would be known), but I believe that one can argue further and refute this possibility. 
The issue is that a neutral pion consists of $q \bar{q}$ pairs. Evidently, we descended one level into the structure of matter, but the question still remains: What are the EM contributions to the quark `physical' mass? 
Therefore, it makes sense to expect that the hadronic mass of the neutral pion should be smaller than its physical mass. At the end of the day, the hadronic mass of the nucleon is unknown, but should be smaller than the 
physical mass of the proton; the hadronic mass of the pion is unknown, but should be smaller than the physical mass of the neutral pion. We must agree on something before attempting a solution to this problem: Are the EM 
corrections supposed to also remove the EM contributions to the physical masses, so that the particles could interact via their hadronic masses in the hypothetical Universe?

The current correction schemes, used in the removal of the EM contributions, assume that the proton in this Universe and the proton in the hypothetical Universe have the same mass, namely the physical mass of the proton. 
The same applies to the other particles, i.e., to the neutron and to the pions. This is the reason that, since 2006, the authors of Ref.~\cite{matsinos2006} have distinguished between stage-1 and stage-2 EM corrections. The 
stage-1 corrections provide estimates for the effects of the Coulomb interaction and, in the case of $\pi^- p$ scattering, for the external mass differences and for the $\gamma n$ channel. Assumed in the derivation of the 
stage-1 corrections was that the hadronic masses of the proton and of the charged pion are equal to their physical masses. The stage-2 EM corrections go one step further; they should take account of graphs with internal 
photon lines, as well as of the effects relating to the use of the physical masses of the particles in the stage-1 corrections, instead of the hadronic ones.

The three works of the Aarhus-Canberra-Zurich Collaboration on the EM corrections in the $\pi N$ system aimed at the removal of the stage-1 effects in low-energy scattering \cite{gashi2001a,gashi2001b}, as well as at 
threshold \cite{oades2007}, \emph{in a consistent manner}. If the EM corrections of Refs.~\cite{gashi2001a,gashi2001b} are applied to the $\pi N$ scattering data, it would be inconsistent to apply corrections to the 
scattering lengths $a_{cc}$ and $a_{c0}$ other than those extracted in Ref.~\cite{oades2007}. The application of another correction scheme would automatically invalidate any comparison between the corrected values of the 
scattering lengths (extracted from the measurements at threshold) and those obtained on the basis of an extrapolation from the scattering data. I believe that the treatment of the stage-2 corrections is well beyond the 
capability of a simple potential model.

I think that (even if they could be conducted somewhere) new $\pi N$ experiments at low energy would not bring much betterment in our knowledge. An advancement of knowledge in low-energy Pion Physics could only be 
instigated by a theoretical breakthrough, in particular in relation to the reliable removal of the EM effects from the various $\pi N$ scattering amplitudes. I have my doubts that ChPT is a promising place to look for such 
a breakthrough. There might be more hope in a non-perturbative approach, such as in Lattice QCD.

\clearpage
\newpage
\section{\label{App:AppB}On obtaining an estimate for $f_c^2$ using the Goldberger-Miyazawa-Oehme (GMO) sum rule}

The GMO sum rule relates the isovector hadronic scattering length $\tilde{b}_1$ with $f_c^2$ \cite{goldberger1955}. The relation reads as
\begin{equation*}
f_c^2 = - \frac{1}{2} \left( 1 - \left( \frac{m_c}{2 m_p} \right)^2 \right) \left( \frac{m_c^2 J^-}{(\hbar c)^2} + \left( 1 + \frac{m_c}{m_p} \right) \frac{m_c \tilde{b}_1}{\hbar c} \right) \, \, \, ,
\end{equation*}
where $J^-$ is defined as
\begin{equation} \label{eq:EQ14}
J^- = \frac{1}{4 \pi^2} \int_0^\infty \frac{ \sigma_{\pi^- p}^T (q) - \sigma_{\pi^+ p}^T (q) }{ \sqrt{q^2 + m_c^2} } dq \, \, \, ;
\end{equation}
the quantities $\sigma_{\pi^\pm p}^T (q)$ denote total cross sections, not containing any EM contributions. Recent estimates for $J^-$ are given in Table \ref{tab:JMinus}. A weighted average was obtained using only the 
statistical uncertainties of the three entries of this table, and the statistical uncertainty of this average was corrected for the quality of the fit via the application of the Birge factor. An average systematic 
uncertainty was obtained from Refs.~\cite{ericson2002,abaev2007} and was quadratically combined with the statistical uncertainty of the weighted average. The $J^-$ value, thus obtained, is equal to $-1.059(32)$ mb; this 
value is used for the extraction of $f_c^2$ estimates in Table \ref{tab:EMCorrections}. The large quoted uncertainty of $J^-$ reflects the magnitude of the systematic effects reported in Refs.~\cite{ericson2002,abaev2007}. 
(The early determinations of $J^-$ from the Karlsruhe analyses, not quoted in this work, are consistent with the values reported in Table \ref{tab:JMinus}, see Refs.~\cite{gibbs1998,ericson2002,abaev2007} for details.)

\begin{table}
{\bf \caption{\label{tab:JMinus}}}Recent estimates for the integral $J^-$ of Eq.~(\ref{eq:EQ14}). All values are expressed in mb. A systematic uncertainty has not been reported in Ref.~\cite{gibbs1998}.
\vspace{0.3cm}
\begin{center}
\begin{tabular}{|c|c|c|c|c|c|}
\hline
Source & $J^-$ & $\delta J^-$ (stat.) & $\delta J^-$ (syst.)\\
\hline
\hline
\cite{gibbs1998} & $-1.051$ & $0.005$ & $-$ \\
\cite{ericson2002} & $-1.083$ & $0.009$ & $0.031$ \\
\cite{abaev2007} & $-1.060$ & $0.007$ & $0.030$ \\
\hline
\end{tabular}
\end{center}
\vspace{0.5cm}
\end{table}

\end{document}